\documentclass{article} 
\usepackage{iclr2020_conference,times}

\usepackage{amsmath,amsfonts,bm}

\def\eqref#1{equation~\ref{#1}}

\def\1{\bm{1}}

\DeclareMathAlphabet{\mathsfit}{\encodingdefault}{\sfdefault}{m}{sl}
\SetMathAlphabet{\mathsfit}{bold}{\encodingdefault}{\sfdefault}{bx}{n}

\usepackage{hyperref}
\usepackage{url}
\usepackage{lineno}
\usepackage{algorithm}
\usepackage{algorithmicx}
\usepackage{subfigure}
\usepackage{amsmath}
\usepackage{graphicx}
\usepackage{multirow}
\usepackage{cases}
\usepackage{subeqnarray}
\usepackage{xspace}
\usepackage{xcolor}
\usepackage{amsfonts}
\usepackage{amsthm}
\usepackage[noend]{algpseudocode}
\usepackage{diagbox}
\usepackage{enumitem}
\usepackage{booktabs}
\modulolinenumbers[5]

\newcommand{\method}{SOT\xspace}
\newcommand{\methodda}{SSDA\xspace}
\newcommand{\equationname}{Equation}
\newcommand{\algorithmname}{Algorithm}

\title{Cross-domain Activity Recognition via Substructural Optimal Transport}

\author{
	$\mathrm{Wang~~Lu}^{\ddagger,\P}$,~~$\mathrm{Yiqiang~~Chen}^{\ddagger,\P}$\thanks{Corresponding Author},~~$\mathrm{Jindong~~Wang}^{\S}$,~~$\mathrm{Xin~~Qin}^{\ddagger,\P}$\thanks{Email: \{luwang, yqchen, qinxin18b\}@ict.ac.cn, Jindong.Wang@microsoft.com}\\
	$\ddagger$ Institute of Computing Technology, Chinese Academy of Sciences, 100190, Beijing, China\\
	$\P$ University of Chinese Academy of Sciences, 100190, Beijing, China\\
	$\S$ Microsoft Research Asia, Beijing, China
}

\begin{document}

\maketitle

\begin{abstract}
	It is expensive and time-consuming to collect sufficient labeled data for human activity recognition (HAR). Domain adaptation is a promising approach for cross-domain activity recognition. Existing methods mainly focus on adapting cross-domain representations via domain-level, class-level, or sample-level distribution matching. However, they might fail to capture the fine-grained locality information in activity data. The domain- and class-level matching are too coarse that may result in under-adaptation, while sample-level matching may be affected by the noise seriously and eventually cause over-adaptation. In this paper, we propose substructure-level matching for domain adaptation (\methodda) to better utilize the locality information of activity data for accurate and efficient knowledge transfer. Based on \methodda, we propose an optimal transport-based implementation, Substructural Optimal Transport (\method), for cross-domain HAR. We obtain the substructures of activities via clustering methods and seeks the coupling of the weighted substructures between different domains. We conduct comprehensive experiments on four public activity recognition datasets (i.e. UCI-DSADS, UCI-HAR, USC-HAD, PAMAP2), which demonstrates that \method significantly outperforms other state-of-the-art methods w.r.t classification accuracy ($\mathbf{9}\%+$ improvement). In addition, \method is $\mathbf{5} \times$ faster than traditional OT-based DA methods with the same hyper-parameters. 
\end{abstract}

\section{Introduction}\label{sec:intro}

Human activity recognition (HAR) plays an important role in ubiquitous computing. Through collected raw signals from sensors, it can easily learn high-level knowledge about human activity. HAR has wide applications in many areas such as gait analysis~\citep{zhao2018hybrid}, gesture recognition~\citep{jia2020classification} and sleep stage detection~\citep{zhao2017learning}. The success of HAR is dependent on an accurate and robust machine learning model. However, to build such good models, we always need to acquire sufficient labeled training data which is time-consuming and expensive. To build models for a new activity dataset that has extremely few or even no labels, a promising approach is to transfer the knowledge learned on the labeled activity data from an \emph{auxiliary} dataset that is similar to this target dataset. The new problem is referred to as the cross-domain activity recognition (CDAR)~\citep{wang2018stratified} since we aim to build models for the target domain (dataset) by leveraging knowledge from the source domain (auxiliary dataset).

It is not appropriate to use labeled activity data from an auxiliary dataset directly, due to different distributions between the auxiliary and the target datasets. Domain adaptation (DA)~\citep{pan2009survey} is a popular paradigm to bridge the distribution gap between two domains for knowledge transfer.
Thus, its key is to match the cross-domain distributions.
A fruitful line of work~\citep{khan2018scaling,rokni2018autonomous} on DA based HAR has achieved great success. We divide these work into two categories according to their different distribution matching schemes. The first category is rough matching which includes the domain-level matching methods~\citep{fernando2013unsupervised,sun2015return}, the class-level matching methods~\citep{zhu2020deep,wang2018stratified} and both domain- and class-level matching methods~\citep{wang2018visual}. These works try to match distributions by learning domain-invariant representations, class-invariant representations, or invariant distributions in both domain and class. The other category is sample-level matching such as Optimal Transport for Domain Adaptation (OTDA)~\citep{courty2016optimal} and Hypergraph Matching based Domain Adaptation~\citep{das2018unsupervised,das2018sample}. The core of these works is to achieve pair-wise sample alignment for two domains.

\begin{figure}[t!]
\centering
\includegraphics[width=\textwidth]{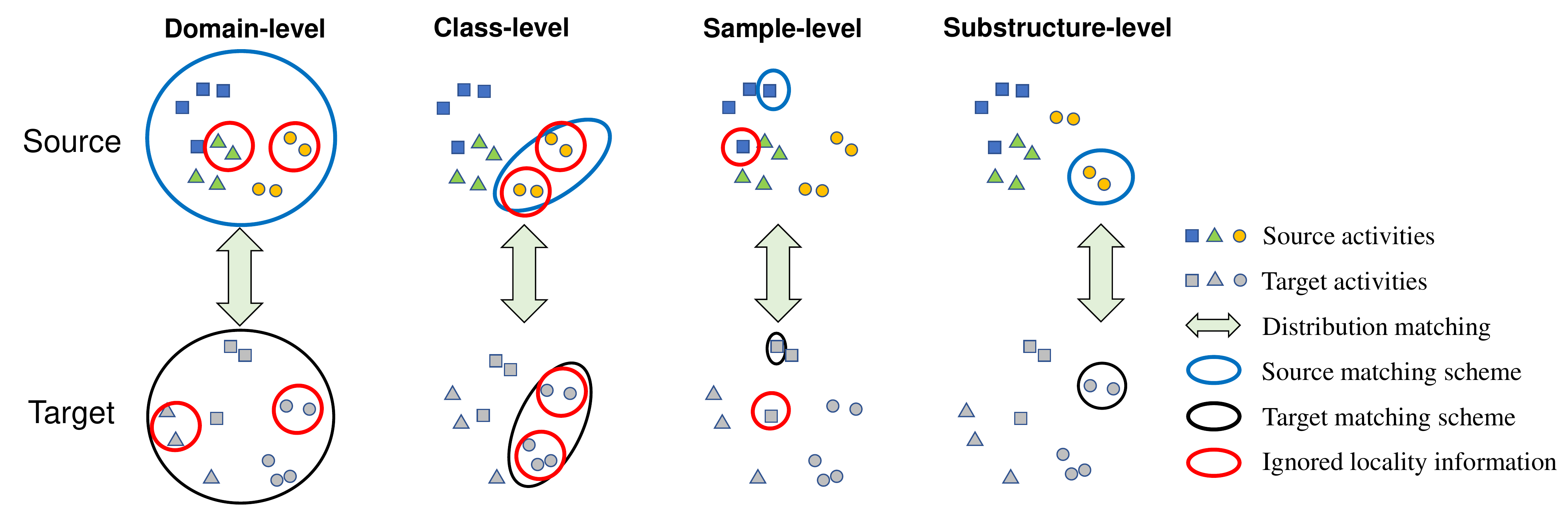}
\vspace{-.2in}
\caption{Comparison of different distribution matching schemes in domain adaptation.}
\vspace{-.2in}
\label{fig:diff-level}
\end{figure}

For CDAR problems, we argue that these methods can suffer from the \emph{under-adaptation} and \emph{over-adaptation} issues since they may ignore the locality information contained in activity data. The locality refers to the fine-grained similarity between two sensor signals that should be considered for domain adaptation. As shown in \figurename~\ref{fig:diff-level}, domain-level matching completely ignores the intra-domain data structure while class-level matching takes a slightly finer alignment~\citep{wang2018stratified}. However, we find that there exist two un-adapted clusters in activity 2 which means rough matching may result in under-adaptation, i.e., domain- and class-level matching fails to capture the locality information. On the other hand, it is obvious that sample-level methods may be seriously affected by some bad points, such as noisy points or outliers, resulting in over-adaptation, i.e., sample-level matching suffers from overfitting when learning the locality information. In addition, sample-level methods need to match too many points, which is notoriously time-consuming. 

In this paper, we tackle domain adaptation based HAR from a different perspective, and hence propose \textbf{Substructural Domain Adaptation (\methodda)} for accurate and robust domain adaptation. Generally speaking, \emph{Substructure} describes the fine-grained latent distribution of data. For the entire class, it can be understood as a data cluster of the class. \figurename~\ref{fig:diff-level} shows how \methodda performs \emph{substructure-level} matching. Compared with the domain- and class-level methods, \methodda utilizes more fine-grained locality information to overcome under-adaptation.
In contrast to sample-level methods that can be easily affected by noise or outliers, \methodda utilizes the substructure of the data that can prevent over-adaptation.
Based on \methodda, we propose an optimal transport-based implementation, namely, \emph{Substructural Optimal Transport (\method)}, for the cross-domain HAR problems.
OT~\citep{villani2008optimal} has a solid theoretical background and allows a flexible mapping without being restricted to a particular hypothesis class~\citep{redko2017theoretical}, hence it has no restriction on the numbers of substructures in different domains. 
Specifically, \method first obtains the internal substructures by a clustering method. Then, the activity substructure weights of the source domain are given via partial optimal transport according to its distance from the target domain.
Finally, two representations of substructures are used to learn a transportation plan matching the probability distribution functions (PDFs) on both substructures.
We conduct experiments on four public activity recognition datasets (UCI DSADS~\citep{barshan2014recognizing}, UCI-HAR~\citep{anguita2012human}, USC-HAD~\citep{mi12:ubicomp-sagaware}, and PAMAP2~\citep{reiss2012introducing}). The results demonstrate that our method outperforms other state-of-the-art methods with a significant improvement of over $\mathbf{9}\%$ w.r.t classification accuracy, while it is $\mathbf{5} \times$ faster than traditional OT-based DA methods with the same hyper-parameters.

Our contributions are mainly three-fold:

\begin{enumerate}
\item We propose \methodda for accurate and robust domain adaptation. \methodda can overcome the under-adaptation and over-adaptation issues in existing domain-, class-, or sample-level matching methods.
\item Based on \methodda, we propose an optimal transport-based implementation, \method, to perform cross-domain activity recognition.
\item Comprehensive experiments on four public activity recognition datasets demonstrate the superiority of \method ($\mathbf{9}\%+$ improvement in accuracy). In addition, \method is $\mathbf{5} \times$ faster than traditional OT-based DA methods with the same hyper-parameters.
\end{enumerate}


\section{Related Work}\label{sec:relate}
\subsection{Human Activity Recognition}\label{subs:r-har}
Human activity recognition has been a popular research topic in ubiquitous computing for its important role in human daily life. HAR attempts to identify and analyze human activities using learned high-level knowledge from raw data of various types of sensors. Several surveys have summarized recent progress in HAR.~\citep{bux2017vision,beddiar2020vision} summed up vision-based HAR and~\citep{wang2019deep,dang2020sensor} summarized recent work from a different point. In this paper, we focus on work about sensor-based HAR. 

Conventional machine learning approaches often treat HAR as a standard time series classification problem. After getting raw data from sensors, they first preprocess the raw data which includes denoising~\citep{castro2016all} and segmentation~\citep{triboan2019semantics}. Then feature extraction and feature selection~\citep{dawn2016comprehensive} are implemented to extract useful features from the pre-processed data. In the following, different models, such as random forest (RF)~\citep{hu2018novel}, Bayesian networks~\citep{xiao2018action}, and support vector machine (SVM)~\citep{reyes2016transition,chen2018infrared}, are built with selected features. Finally, we can use these learned models to make activity inferences. Deep learning based HAR~\citep{ignatov2018real,zhao2018deep,khan2017detecting,hassan2018human,hassan2018robust} automatically extracts abstract features through several hidden layers and reduces the effort of choosing the right features.

However, most of the methods mentioned before assume that the training and testing data are in the same distribution, which is not suitable for CDAR problems. As for CDAR problems, source and target data are usually from a different distribution, which results in weak generalization of the aforementioned methods. Therefore, approaches to the CDAR are needed. And in this paper, we mainly focus on the traditional approaches for CDAR problems.

\subsection{Transfer Learning and Domain Adaptation}\label{subs:r-tlda}
Transfer learning tries to leverage source domain knowledge to help learn models in the target domain, which mitigates the problem that the target domain has no label or few labels.~\citep{pan2009survey, weiss2016survey} concluded the traditional transfer learning methods, and~\citep{tan2018survey,wilson2020survey} introduced the deep transfer learning methods. Domain adaptation, as a branch of the transfer learning, solves the problem that a distribution shift exists between different domains and has been successfully applied in many applications, such as visual image classification~\citep{wang2018deep}, natural language processing~\citep{li2020unsupervised}, sentiment classification~\citep{dai2020adversarial}, etc. We roughly divide domain adaptation into two categories: 1) rough matching which includes domain-level matching, class-level matching, and both domain- and class-level matching; 2) sample-level matching which is a meticulous matching way.

Domain adaptation has developed for many years and most of those methods exploit rough matching between two domains.~\citep{pan2010domain} proposed a method, named Transfer Component Analysis (TCA), which learns a kernel in the reproducing kernel Hilbert space (RKHS) to minimize the maximum mean discrepancy (MMD) between domains. \citep{wang2018stratified} proposed stratified transfer learning (STL) and achieved the goal of intra-class transfer. Joint
distribution adaptation (JDA)~\citep{long2013transfer} is based on minimizing joint distribution between domains.~\citep{wang2017balanced,wang2020transfer} extended it and proposed Balanced Distribution Adaptation (BDA) which adaptively adjusts the importance of marginal distribution and conditional distribution. \citep{zhao2020local}
proposed a method, named Local Domain Adaptation (LDA), which takes a compromise between domain- and class-level matching and utilizes high-level abstract clusters to organize data.

Sample-level matching is the other way to match distribution between domains and the representative work is~\citep{courty2016optimal}. Nicolas Courty et al. utilized the theory of optimal transport to learn the coupling between two probability density functions. Through the barycentric mapping~\citep{villani2008optimal}, the images of the source samples in the target domain are obtained, and then a simple classification model can be used to classify the target samples.~\citep{kerdoncuff2020metric} extended it which designs a metric learning optimal transport (MLOT) algorithm to optimizes a mahalanobis distance. Besides optimal transport based methods,~\citep{das2018unsupervised,das2018sample} used hypergraph matching to match the samples between two domains.

\methodda is different from these methods. It takes advantage of the substructure of domains and utilizes the substructure-level matching to seek the balance of rough matching and sample-level matching.

\subsection{Human Activity Recognition with Transfer Learning}\label{subs:r-hartl}
There is much prior work focusing on HAR with transfer learning and a detailed survey can be found in~\citep{cook2013transfer}.

\citep{zhao2011cross} proposed an algorithm known as transfer learning embedded decision tree (TransEMDT) which integrates a decision tree and the k-means clustering algorithm to solve the cross-people activity recognition problem. Lin et al.~\citep{lin2016cross} identified a compact joint subspace for each class and then measured the distance between classes using principal angle.~\citep{wang2018stratified} tried to learn more reliable pseudo labels using the majority voting technique on both domains.~\citep{rey2017label} considered a special case that the new domain just contains the old one and~\citep{feuz2017collegial} proposed a heterogeneous transfer learning method for HAR. Recently, Qin et al.~\citep{qin2019cross}  proposed an adaptive spatial-temporal transfer learning (ASTTL) approach to select the most similar source domain to the target domain and accurately transfer activity. Despite many approaches have been designed to solve the CDAR problem and some of them attempted to use clustering methods, little work is substructure-based. \method tries to complete substructure-level matching through joint Gaussian Mixture Model and optimal transport. The number of substructures may be different from the number of the classes.

\section{Method}\label{sec:method}

\subsection{Problem Formulation}\label{subs:m-problem}
In a CDAR problem~\citep{wang2018stratified}, a labeled source domain $\mathcal{D}_s = \{(\mathbf{x}_i, y_i)\}_{i=1}^{n_s}$ and an unlabeled target domain $\mathcal{D}_t = \{\mathbf{x}_j \}_{j=1}^{n_t}$ are given, where $n_s$ and $n_t$ are the number of source and target samples respectively. In our problem, $\mathcal{D}_s$ and $\mathcal{D}_t$ have the same feature spaces and label spaces, i.e. $\mathcal{X}_s = \mathcal{X}_t \subset \mathbb{R}^d$ and $\mathcal{Y}_s = \mathcal{Y}_t$, where $d$ is the feature dimension. $y_s, y_t \in \{1, \cdots, C \}$, and $C$ is the number of classes. Two domains have different distributions, i.e., $p_s(\mathbf{x},y) \ne p_t(\mathbf{x}, y)$. The goal of cross-domain learning is to obtain the labels $y_t$ for the target domain with the help of the source domain $\mathcal{D}_s$. 

\subsection{Motivation}\label{subs:m-moti}
In a CDAR problem, labeled source data often has a different distribution with target data. For example, source data may be collected from the sensor tied on the front right hip while target data may contain waist-mounted smartphone data.
When we perform rough distribution matching which aligns whole data or aligns data based on classes, we may not be able to match data perfectly since different people have their styles even performing the same activity. Thus, we need to capture the locality information.

The raw activity data can be represented as $\mathbf{x} = \mathbf{z} + \boldsymbol{\delta}$, where $\mathbf{z}\in \mathbb{R}^d$ represents an activity prototype containing the data collected from the standard activity in an ideal situation and $\boldsymbol{\delta}$ corresponds to the noise in reality.
Therefore, sample-level matching that aligns $\mathbf{x}$ directly may introduce noise, and performing sample-level matching might have no practical meaning. Overall, a compromise between rough matching and sample-level matching is needed to obtain more fine-grained alignments and avoid noise influence. 

Apart from empirical analysis, we theoretically analyze our motivation by formulating the distribution as: 
\begin{subequations}
\label{eqa:moti}
\begin{align}
p(\mathbf{x}) &=\sum_y p(\mathbf{x}|y)p(y)\label{eqa:class}\\
&=\sum_y(\sum_o p(\mathbf{x},o|y))p(y)\notag\\
&=\sum_y\sum_o p(\mathbf{x}|y,o)p(y,o) \text{~(For source domain)}\label{eqa:sproto} \\
&=\sum_o\sum_y p(\mathbf{x}|y,o)p(y|o)p(o)\notag\\
&=\sum_o p(\mathbf{x}|o)p(o). \text{~(For target domain)}\label{eqa:tproto} 
\end{align}
\end{subequations}

According to \equationname~\eqref{eqa:class}, domain-level matching tries to match $p(\mathbf{x})$ while class-level matching tries to match $p(\mathbf{x}|y)$. From the previous analysis, we know that one class may include more fine-grained locality information, which means class-level matching may not be enough. Therefore, a deeper decomposition of $\mathbf{x}$ is needed. We denote $o$ as the locality information contained in each $y$, i.e., $o$ is the \emph{substructure}. As shown in \equationname~\eqref{eqa:sproto}, we can divide the labeled source domain into multiple substructures. Since the target domain has no labels, further conversion is performed and \equationname~\eqref{eqa:tproto} shows we can divide the target domain into finer substructures.

We know that 
\begin{equation}
p(y|o)=
\begin{cases}
1& o \mathrm{~~ is~~ part~~ of~~ } y \\
0& o.w. \\
\end{cases}
\label{eqa:zy}
\end{equation}
Denote the substructure of $o$ as $y_o$, then $\sum_y \sum_o  p(\mathbf{x}|y, o) p(y, o) = \sum_o p(\mathbf{x}, y_o|o) p(o)$, which indicates that \equationname~\eqref{eqa:sproto} and \equationname~\eqref{eqa:tproto} are identical. Now, we can match $p(\mathbf{x}|o)$. Obviously, this substructure-level matching is more fine-grained compared with domain- and class-level matching while it avoids the influence of noise via the use of substructures compared with sample-level matching. 

\tablename~\ref{tab:matchinginfo} compares different matching schemes.

\begin{table}[htbp]
\caption{Comparison between different matching schemes.}
\centering
\label{tab:matchinginfo}
\resizebox{\textwidth}{!}{
	\begin{tabular}{c c c c}
		\toprule
		\textbf{Type}&\textbf{Assumption}&\textbf{Formulation}&\textbf{Limitations}\\
		\hline
		Domain-level&domain-invariant features&$p(\mathbf{x}_s)\longleftrightarrow p(\mathbf{x}_t)$&too coarse matching\\
		
		Class-level&class-invariant features&$p(\mathbf{x}_s|y_s)\longleftrightarrow p(\mathbf{x}_t|y_t)$&coarse matching\\
		
		Substructure-level&substructure-invariant features &$p(\mathbf{x}_s,y_s|o_s)\longleftrightarrow p(\mathbf{x}_t|o_t)$&\diagbox{}{}\\
		
		Sample-level&no strict restrictions&$(\mathbf{x}_s,y_s)\longleftrightarrow \mathbf{x}_t$& affected by noise, low-efficiency \\
		\bottomrule
	\end{tabular}
}
\end{table}

\begin{figure}[htbp]
\centering
\subfigure[toy dataset distribuion]{
	\label{fig:toy data}
	\includegraphics[width=0.3\textwidth]{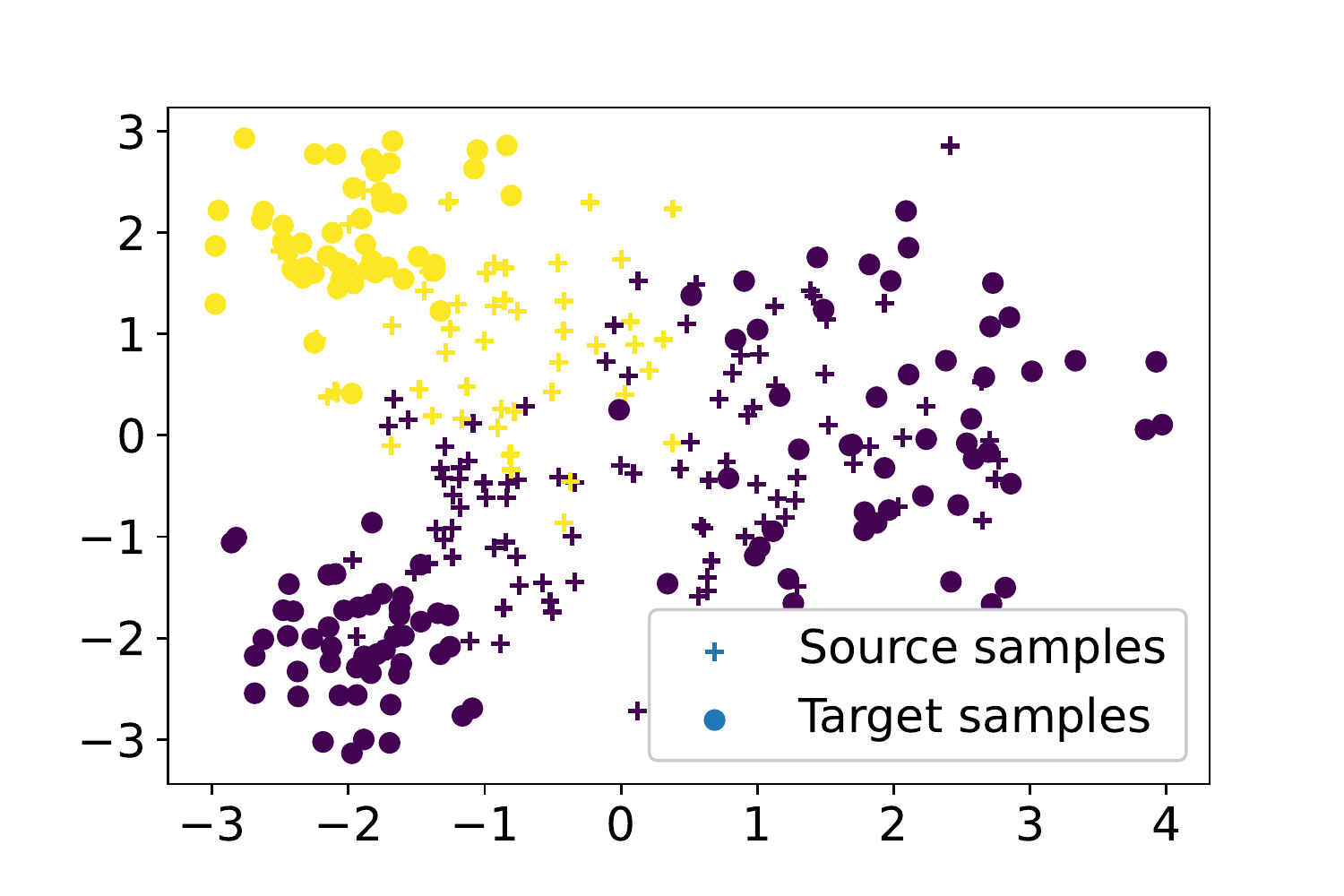}
}
\subfigure[OTDA for toy dataset]{
	\label{fig:OTDA_syn}
	\includegraphics[width=0.3\textwidth]{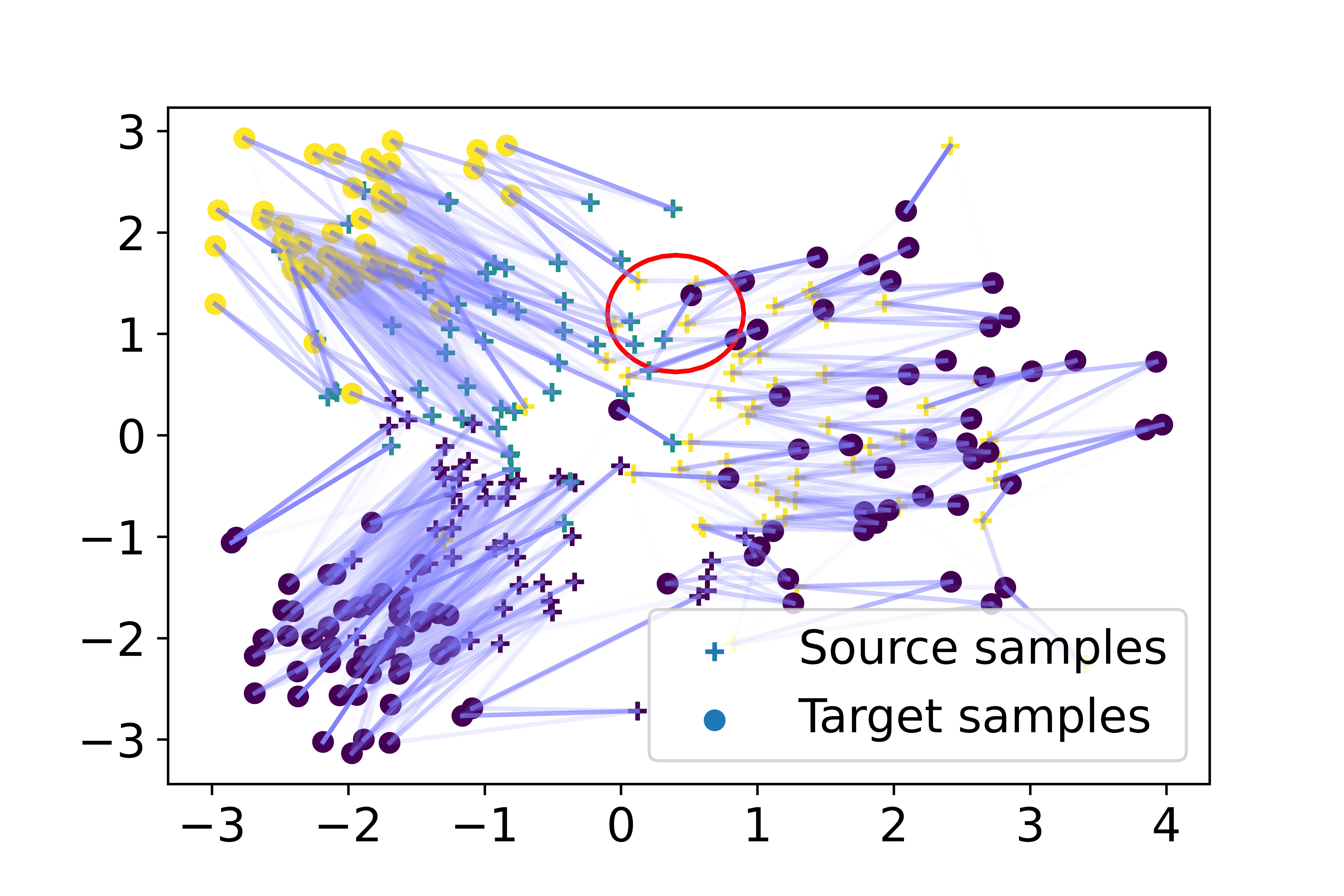}
}
\subfigure[\method for toy dataset]{
	\label{fig:gmmot_syn}
	\includegraphics[width=0.3\textwidth]{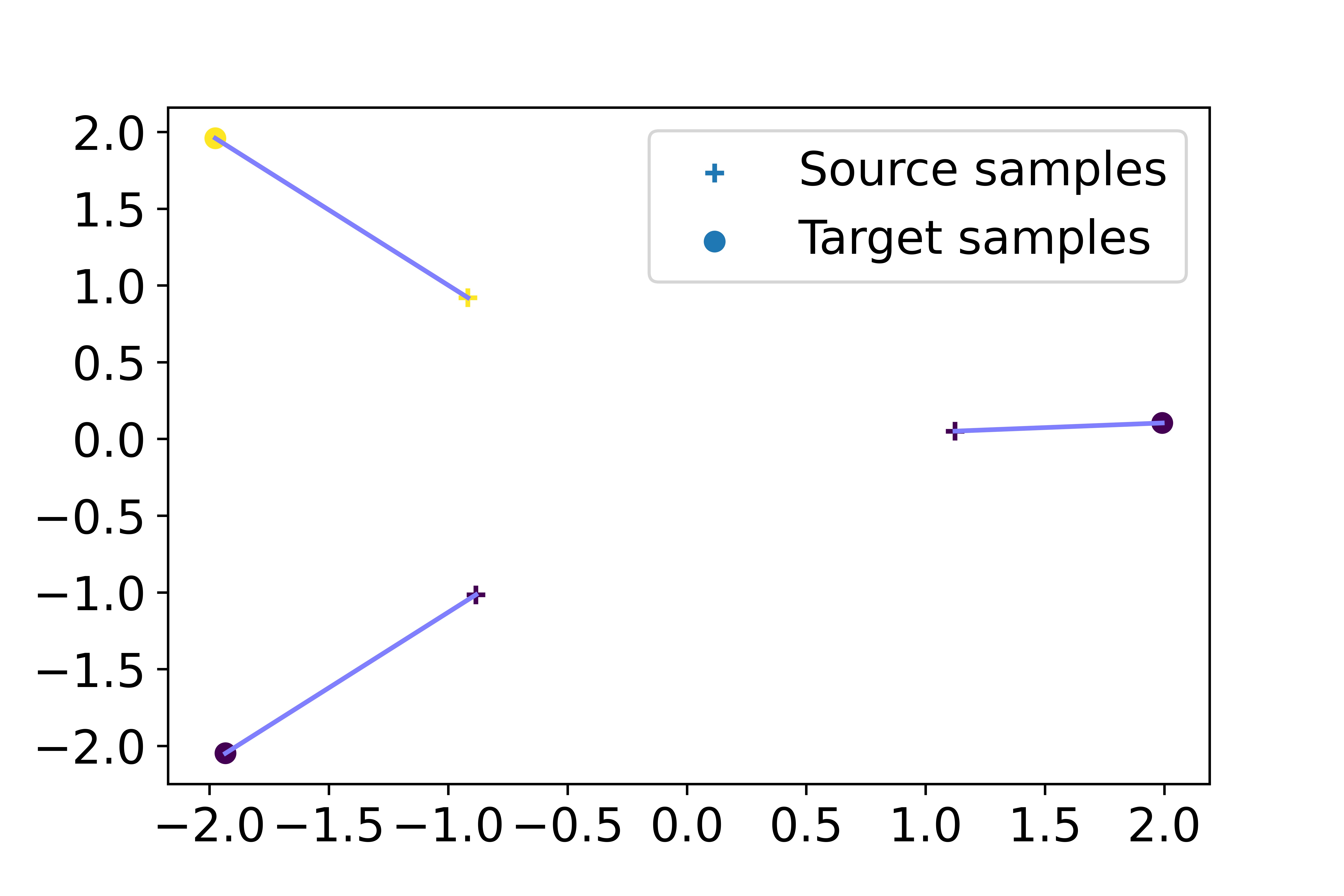}
}
\caption{Toy example to show the effectiveness of substructure-level matching.}
\label{fig:syn}
\end{figure}

To explain the necessity of using the substructures more clearly, we give a toy example. As we can see from \figurename~\ref{fig:toy data}, the source has three clusters that are sampled from a Gaussian Mixture Model (GMM) and the target also has three clusters sampled from a slightly different GMM. Both domains have two classes while the different colors respond to the different classes. It is obvious that one of the classes has two components, which means rough matching may not be suitable. Next, we consider the sample-level matching. The concrete data can be treated as the noisy version of the prototypes, i.e. the cluster centers adding perturbation. \figurename~\ref{fig:OTDA_syn} shows that if we directly match two domains with concrete data points, there will be miss matching. The red circle in \figurename~\ref{fig:OTDA_syn} points one miss matching. Intuitively speaking, the matching of noise points to noise points has no practical meaning.

\begin{figure}[t!]
\centering
\includegraphics[width=\textwidth]{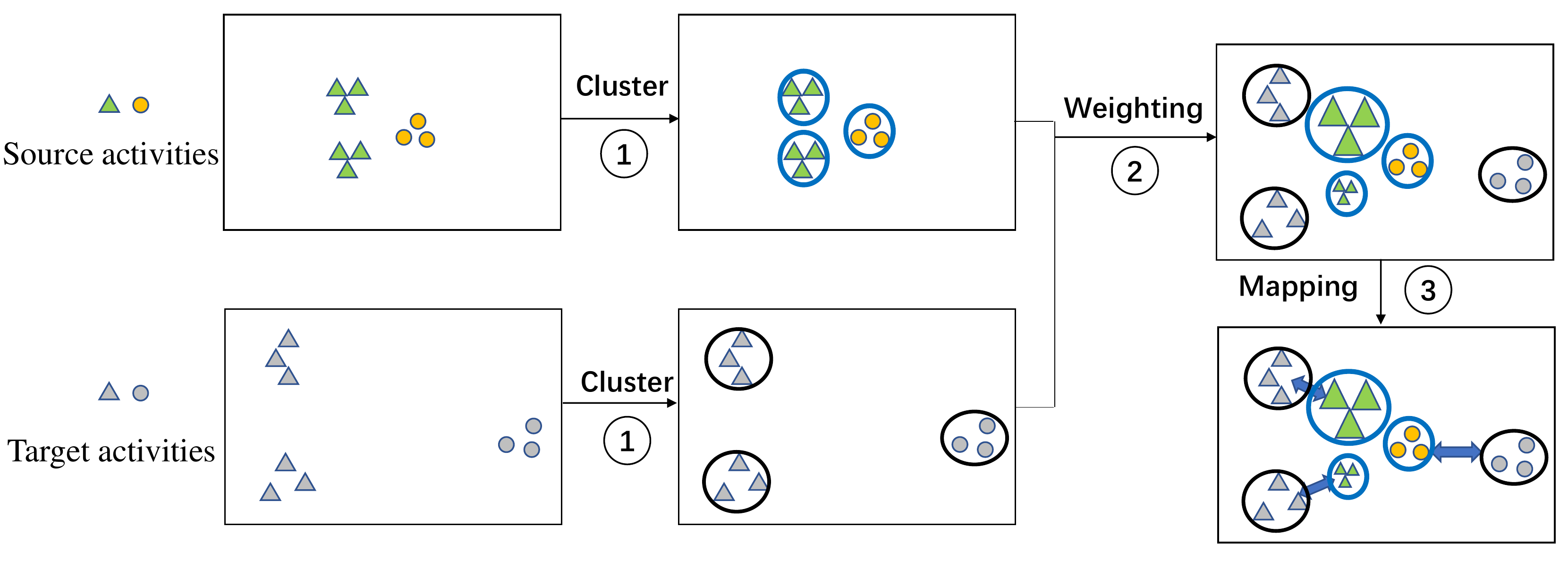}
\vspace{-.2in}
\caption{Overview of the \methodda framework.}
\label{fig:frame}
\end{figure}

\subsection{\methodda: A general framework for domain adaptation}\label{sub-SSDA}
In this paper, we propose a general \textbf{Substructural Domain Adaptation (\methodda)} method.
\methodda is a general framework consistent with the substructure and \figurename~\ref{fig:frame} illustrates the main process of \methodda which mainly contains three steps. Firstly, \methodda clusters the data to obtain the substructures of activities. As a general framework, we can choose a suitable clustering algorithm for customization. Then, it gives weights to the source substructures according to priors or adaptive methods. Weight represents the importance of substructures and different substructures often play different roles. For example, some substructures far away from most data should play small roles with small weights. For simplicity, we usually give uniform weights to all structures without priors. Finally, mapping is performed on the substructures of different domains.
We can extend some traditional method, such as CORrelation alignment (CORAL)~\citep{sun2015return}, to perform substructure-level matching, which is really commendable.

\subsection{\method: an OT implementation of \methodda}\label{sub-SOT}

In this section, we propose an OT implementation of \methodda, named \method.
\method utilizes GMM to get substructures while it uses OT to perform weighting and mapping. According to the substructure representation, we introduce $\mathrm{\method}_c$ with center representation and $\mathrm{\method}_g$ with distribution representation.

\subsubsection{Substructures Generation And Representations}\label{subs:m-sgar}
We denote $\boldsymbol{\delta} \sim \mathcal{N}(0; \boldsymbol{\sigma}^2)$ and $\mathbf{X}$ represents all feature data. Equivalently, $\mathbf{X}_k$ conforms to a Gaussian distribution whose center is the corresponding prototypes, i.e. $\mathbf{X}_k \sim \mathcal{N}(\mathbf{z}_k, \boldsymbol{\sigma}_k)$. 
$\mathbf{z}_k$ means the value of $k$th center, $\boldsymbol{\sigma}_k$ means the $k$th covariance, and $\mathbf{X}_k$ means the data belong to the $k$th cluster. Now, we have data $\mathbf{X}$ and our goal is to get $\mathbf{z}_k$ and $\boldsymbol{\sigma}_k$. It is easy to use Expectation  Maximum (EM)~\citep{dempster1977maximum} algorithm to obtain the parameters of the Gaussian Mixture Models. 

To maintain label consistency in the source domain, we treat the source domains as a mixture distribution of $C$ Gaussian mixture models and each one corresponds to one class in the source domain. The number of components is determined by the Bayesian Information Criterion (BIC)~\citep{schwarz1978estimating}, i.e. 
\begin{equation}
\label{eqa:BIC}
BIC = -2 \ln (L) + k\ln(m),
\end{equation}
where $L$ represents the maximized value of the likelihood function for the estimated model, $k$ represents the number of free parameters to be estimated, and $m$ is the sample size. We seek $K$ which minimizes BIC. Due to the lack of labels in the target domain, we have to perform clustering on the entire target domain and the number of clusters is up to the specific dataset.

After getting the clusters in the source domain and the target domain, we design two different ways to represent the substructures which correspond to $\mathrm{\method}_c$ and $\mathrm{\method}_g$ respectively. $\mathrm{\method}_c$ with center representation utilizes only information from cluster center, and it is simple and computationally efficient while $\mathrm{\method}_g$ with distribution considers more information on clusters, but it needs some approximations when computing.  

$\mathrm{\method}_c$:
After clustering, two domains are expressed as the cluster centers, i.e. $\mu_{c,s} = \sum_{i=1}^{k_s} w_{s,i} \delta_{\mathbf{z}_{s,i}}, \mu_{c,t} = \sum_{i=1}^{k_t} w_{t,i} \delta_{\mathbf{z}_{t,i}}$. $\mathbf{z}\in \mathbb{R}^d$ represents the cluster centers and $\delta_{\mathbf{z}}$ is a Dirac function at location $\mathbf{z}$. $\mu_{c,s}, \mu_{c,t}$ are distributions of the source domain and the target domain respectively and $w$ are probability masses associated to the $\mathbf{z}$. Obviously, $\sum_{i=1}^{k_s} w_{s,i} = 1, \sum_{i=1}^{k_t} w_{t,i} = 1$. In addition, the squared Euclidean distance can be chosen as the cost between $\mathbf{z}_{s,i}$ and $\mathbf{z}_{t,j}$, i.e. 
\begin{equation}
\label{eqa:cost1}
c(\mathbf{z}_{s,i}, \mathbf{z}_{t,j}) = ||\mathbf{z}_{s,i} - \mathbf{z}_{t,j}||_2^2.
\end{equation}

$\mathrm{\method}_g$: This one utilizes the cluster distributions to represent the substructures, and the covariance of the clusters can be understood as the difficulty of the activity for the person. Therefore, the source domain can be expressed as $\mu_{g,s} = \sum_{i=1}^{k_s} w_{s,i} \mathcal{N}( \mathbf{z}_{s,i},  \boldsymbol{\sigma}_{s,i})$ and the target domain can be expressed as $\mu_{g,t} = \sum_{i=1}^{k_t} w_{t,i} \mathcal{N}(\mathbf{z}_{t,i}, \boldsymbol{\sigma}_{t,i})$. The meanings of the symbols are similar to the first way. The only difference is that we use a Gaussian distribution $\mathcal{N}(\mathbf{z},\boldsymbol{\sigma})$ instead of a Dirac function $\delta_\mathbf{z}$. In this situation, the squared Wasserstein distance~\citep{peyre2019computational} replaces the squared Euclidean distance as the cost function, i.e.
\begin{equation*}
\begin{aligned}
c(\mathcal{N}(\mathbf{z}_{s,i},\boldsymbol{\sigma}_{s,i}), \mathcal{N}(\mathbf{z}_{t,j},\boldsymbol{\sigma}_{t,j}))& = W_2^2(\mathcal{N}(\mathbf{z}_{s,i},\boldsymbol{\sigma}_{s,i}), \mathcal{N}(\mathbf{z}_{t,j},\boldsymbol{\sigma}_{t,j}))\\ & = ||\mathbf{z}_{s,i} - \mathbf{z}_{t,j}||^2 + B(\boldsymbol{\sigma}_{s,i}, \boldsymbol{\sigma}_{t,j})^2,
\end{aligned}
\end{equation*} 
where $B$ is the Bures metric \citep{bhatia2019bures} between positive definite matrices and can be calculated as follows,
\begin{equation}
B(\boldsymbol{\sigma}_{s,i}, \boldsymbol{\sigma}_{t,j})^2 = tr(\boldsymbol{\sigma}_{s,i} + \boldsymbol{\sigma}_{t,j} - 2(\boldsymbol{\sigma}_{s,i}^{1/2}\boldsymbol{\sigma}_{t,j}\boldsymbol{\sigma}_{s,i}^{1/2} )^{1/2}),
\end{equation}
where $tr(\cdot)$ denotes the trace of a matrix, $\boldsymbol{\sigma}^{1/2}$ is the matrix square root. For simplicity, we force the covariance matrix to be a diagonal matrix, i.e. $\boldsymbol{\sigma} = diag(r_i)_{i=1}^d$. In this case, the Bures metric is the Hellinger distance $B(\boldsymbol{\sigma}_{s,i}, \boldsymbol{\sigma}_{t,j}) = ||\sqrt{\mathbf{r}_{s,i}} - \sqrt{\mathbf{r}_{t,j}}||$. Overall, the cost function is 
\begin{equation}
\label{eqa:cost2}
\begin{aligned}
c(\mathcal{N}(\mathbf{z}_{s,i},\boldsymbol{\sigma}_{s,i}), \mathcal{N}(\mathbf{z}_{t,j},\boldsymbol{\sigma}_{t,j})) &= ||\mathbf{z}_{s,i} - \mathbf{z}_{t,j}||^2 + ||\sqrt{\mathbf{r}_{s,i}} - \sqrt{\mathbf{r}_{t,j}}||_2^2 \\
&= || (\mathbf{z}_{s,i},\sqrt{\mathbf{r}_{s,i}}) - (\mathbf{z}_{t,j},\sqrt{\mathbf{r}_{t,j}})||_2^2.
\end{aligned}
\end{equation}
$\mathbf{r}_{s,i}$ and $\mathbf{r}_{t,j}$ represent diagonals of the $i$th source domain cluster's covariance and the $j$th target domain cluster's covariance respectively. $(\mathbf{z}, \sqrt{\mathbf{r}})$ concatenates the $\mathbf{z}$ and $\sqrt{\mathbf{r}}$ and serves as the new feature of the substructure.

\subsubsection{Weighting Source Substructures }\label{subs:m-wss}
For unity, we denote the source domain as $P_s = \sum_{i=1}^{k_s} w_{s,i} p_{s,i}$ and denotes the target domain as $P_t = \sum_{i=1}^{k_t}w_{t,i} p_{t,i}$. Due to little information about the target domain, we treat $p_{t,i}$ equally and fix $w_{t,i}$ to $1/k_t$. Now, we compute the $w_{s,i}$ adaptively.

Since we only know $\sum_{i=1}^{k_s} w_{s,i} = 1$, it can be seen as a partial optimal transport problem, and the upper bounds of $w_{s,i}$ are all 1. Obviously, the total cost of the partial optimal transport is $\langle\boldsymbol{\pi},\mathbf{C}\rangle_F$, where $\langle\cdot, \cdot\rangle_F$ is the Frobenius dot product, $\mathbf{C}$ is the cost matrix, and $\boldsymbol{\pi}$ is the coupling matrix between two PDFs. For calculation convenience, an entropy item, i.e.$H(\boldsymbol{\pi}) = \sum_{ij}\pi_{ij}\log \pi_{ij}$, is added. Now, our goal is to obtain the optimal transport.
\begin{equation}
\label{eqa:pot1}
\begin{aligned}
\boldsymbol{\pi}^*_1 &= &\arg\min_{\boldsymbol{\pi}} \langle\boldsymbol{\pi}, \mathbf{C}\rangle_F + \lambda_1 H(\boldsymbol{\pi})\\
s.t& &\boldsymbol{\pi}^T \mathbf{1}_{k_s}  =  \mathbf{w}_t\\
&&\boldsymbol{\pi} \mathbf{1}_{k_t} \leq \mathbf{1}_{k_s}\\
&&\mathbf{1}_{k_t}^T\boldsymbol{\pi}^T\mathbf{1}_{k_s} = 1.
\end{aligned}
\end{equation}
$\mathbf{1}_k$ is k-dimensional vector of ones and $\lambda_1$ is a hyper-parameter that balances the calculation speed and accuracy. $H(\boldsymbol{\pi})$ requires $\boldsymbol{\pi} \geq 0$. When $\boldsymbol{\pi} \geq 0$ and $\mathbf{1}_{k_t}^T\boldsymbol{\pi}^T\mathbf{1}_{k_s} = 1$, it is obvious that $\boldsymbol{\pi}\mathbf{1}_{k_t} \leq \mathbf{1}_{k_s}$ always holds. In addition, $\mathbf{1}^T_{k_t} \mathbf{w}_t = 1$ holds, which means $\mathbf{1}_{k_t}^T\boldsymbol{\pi}^T\mathbf{1}_{k_s} = 1$ also always holds. Therefore, \equationname~\eqref{eqa:pot1} can be simplified as the following.
\begin{equation}
\label{eqa:pot2}
\begin{aligned}
\boldsymbol{\pi}^*_1 &= &\arg\min_{\boldsymbol{\pi}} \langle\boldsymbol{\pi}, \mathbf{C}\rangle_F + \lambda_1 H(\boldsymbol{\pi})\\
s.t& &\boldsymbol{\pi}^T \mathbf{1}_{k_s}  =  \mathbf{w}_t.
\end{aligned}
\end{equation}

We denote the feasible solution set of $\boldsymbol{\pi}^T \mathbf{1}_{k_s}  =  \mathbf{w}_t$ as $C_1$. Obviously, the $C_1$ is a convex set. The optimization goal of \equationname~\eqref{eqa:pot2} is also convex. And, it is easy to get the closed form of this problem. In the following, the Lagrange method is adopted to solve the problem.

We denote $\boldsymbol{\phi}$ as the Lagrange multiplier, then our goal can be derived as 
\begin{equation*}
L = \langle\boldsymbol{\pi}, \mathbf{C}\rangle_F + \lambda_1 H(\boldsymbol{\pi}) + \boldsymbol{\phi}^T(\boldsymbol{\pi}^T\mathbf{1}_{k_s} - \mathbf{w}_t).    
\end{equation*}
To get the optimal point, the following equations must hold:
\begin{subequations}
\begin{numcases}{}
	\frac{\partial L}{\partial \boldsymbol{\pi}} = 0\label{eqa:partial}\\
	\boldsymbol{\pi}^T \mathbf{1}_{k_s} - \mathbf{w}_t = \mathbf{0}.\label{eqa:kkt}
\end{numcases}
\label{eqa:lagrange}
\end{subequations}

Using \equationname~\eqref{eqa:partial}, we can get 
$\mathbf{C} + \lambda_1 (1 + \log\boldsymbol{\pi}) + \mathbf{1}_{k_s}\boldsymbol{\phi}^T = \mathbf{0},$ which means
\begin{equation}
\label{eqa:rpart}
\boldsymbol{\pi} = e^{-\frac{\mathbf{1}_{k_s}\boldsymbol{\phi}^T - \mathbf{C}}{\lambda_1} - 1}.
\end{equation}
Then, we substitute \equationname~\eqref{eqa:rpart} into \equationname~\eqref{eqa:kkt} and get
\begin{equation*}
{e^{\frac{-\mathbf{1}_{k_s}\boldsymbol{\phi}^T - \mathbf{C}}{\lambda_1} - 1}}^T \mathbf{1}_{k_s} = \mathbf{w}_t.
\end{equation*}

Therefore, we get
\begin{equation*}
{e^{\frac{ - \mathbf{C}}{\lambda_1} - 1}}^T\odot {e^{\frac{-\mathbf{1}_{k_s}\boldsymbol{\phi}^T}{\lambda_1}}}^T \mathbf{1}_{k_s} = \mathbf{w}_t,
\end{equation*}
where $\odot$ means element-wise product. Obviously, each element in the same row of $\exp(\frac{-\mathbf{1}_{k_s}\boldsymbol{\phi}^T}{\lambda_1})$ is the same number, and we can easily get the optimal $\boldsymbol{\pi}^*$.
We initialize $\boldsymbol{\pi}_0 = \exp(-\frac{\mathbf{C}}{\lambda_1} - 1)$ and get 
\begin{equation}
\label{eqa:rwpot}
\boldsymbol{\pi}^*_1 = \boldsymbol{\pi}_0 \mathrm{diag}(\mathbf{w}_t \oslash \boldsymbol{\pi}_0^T \mathbf{1}_{k_s}),
\end{equation}
where $\oslash$ denotes element-wise divide and $\mathrm{diag}$ denotes diagonals.
Once the optimal coupling matrix $\boldsymbol{\pi}^*_1$ is obtained, the source weight can be easily calculated as $\mathbf{w}_s = \boldsymbol{\pi}^*_1 \mathbf{1}_{k_t}$.

\subsubsection{OT-based Mapping of the Substructures}\label{subs:m-otms}
Through the previous steps, the source domain distribution is $P_s = \sum_{i=1}^{k_s} w_{s,i} p_{s,i}$ while the target domain distribution is $P_t = \sum_{i=1}^{k_t}w_{t,i} p_{t,i}$. And the label corresponding to $p_{s,i}$ is the label of the data belongs to $i$th cluster in the source domain, i.e. $\tilde{y}_{s,i}$. According to \equationname~\eqref{eqa:cost1} or \equationname~\eqref{eqa:cost2}, we can easily get the cost matrix $\mathbf{C}$.
Following~\citep{courty2016optimal}, the objective for SOT is 
\begin{equation}
\label{eqa:gmmot}
\begin{aligned}
\boldsymbol{\pi}^* &= &\arg\min_{\boldsymbol{\pi}} \langle\boldsymbol{\pi}, \mathbf{C}\rangle_F + \lambda H(\boldsymbol{\pi}) + \eta \Omega(\boldsymbol{\pi}) \\
s.t& &\boldsymbol{\pi}^T \mathbf{1}_{k_s}  =  \mathbf{w}_t\\
&&\boldsymbol{\pi} \mathbf{1}_{k_t} = \mathbf{w}_s.
\end{aligned}
\end{equation}
$\Omega(\boldsymbol{\pi})$ is group-sparse regularizer, and it expects that each target sample receives masses only from source samples that have the same label. And following~\citep{courty2016optimal}, we define the regularizer as $\Omega(\boldsymbol{\pi}) = \sum_j \sum_{cl} ||\boldsymbol{\pi}(I_{cl}, j)||_2$, where $||\cdot||_2$ denotes the $l_2$ norm and $I_{cl}$ contains the indices of rows in $\boldsymbol{\pi}$ related to source domain samples of class $cl$. $\boldsymbol{\pi}(I_{cl}, j)$ is a vector containing coefficients of the $j$th column of $\boldsymbol{\pi}$ associated to class $cl$, and it induces the desired sparse representation in the target samples. $\lambda$ and $\eta$ are hyper-parameters.
We use generalized conditional gradient (GCG) following~\citep{courty2016optimal} to solve the optimization problem.

Once obtaining the optimal coupling matrix $\boldsymbol{\pi}^*$, we can compute the transformation of $\mathbf{p}_{s,i}$ by barycentric mapping, i.e. $\hat{\mathbf{p}}_{s,i} = \arg \min_ {\mathbf{p}} \sum_j \pi^*(i,j) c\left(\mathbf{p}, \mathbf{p}_{t,j}\right)$. When the cost function is the squared $l_2$ distance, this barycentric mapping can be
expressed as: 
\begin{equation}
\label{eqa:baymap}
\hat{\mathbf{P}}_s = diag(\boldsymbol{\pi}^* \mathbf{1}_{k_t})^{-1}\boldsymbol{\pi}^*\mathbf{P}_t,
\end{equation}
where $\mathbf{P}_t$ represents the target representation and $\hat{\mathbf{P}}_s$ represents the source mapping representation. Now, any traditional machine learning model, such as 1-Nearest Neighbor (1NN), can be used to learn the predictions with help of $\hat{\mathbf{P}}_s$ and $\tilde{y}_{s,i}$. After getting the label $\tilde{y}_{t,i}$ corresponding to $p_{t,i}$, we assign the same label, $\tilde{y}_{t,i}$, to the data belonging to $p_{t,i}$ in the target domain.

The overall process of \method is described in \algorithmname~\ref{alg:gmmot}.
\begin{algorithm}[t]
\caption{\method: Substructural Optimal Transport} 
\label{alg:gmmot}
\hspace*{0.02in} {\bf Input:} 
source dataset $\mathcal{D}_s = \{(\mathbf{x}_{s,i}, y_{s,i})\}_{i=1}^{n_s}$, target dataset $\mathcal{D}_t = \{ (\mathbf{x}_{t,i}) \}_{i=1}^{n_t}$, hyper-parameters $\lambda_1, \lambda, \eta, k_t$\\
\hspace*{0.02in} {\bf Output:} 
target labels$\{y_{t,i} \}_{i=1}^{n_t}$
\begin{algorithmic}[1]
	\State Use EM for GMM, cluster each class data in the source domain to obtain $\{ (\mathbf{p}_{s,i}, \tilde{y}_{s,i}) \}_{i=1}^{k_s}$. The number of clusters is determined by \equationname~\eqref{eqa:BIC}.
	\State Use EM for GMM to obtain $\{ (\mathbf{p}_{t,i}) \}_{i=1}^{k_t}$
	\State Compute cost matrix $\mathbf{C}$ according \equationname~\eqref{eqa:cost1} ($\mathrm{\method}_c$) or \equationname~\eqref{eqa:cost2} ($\mathrm{\method}_g$)
	\State Use \equationname~\eqref{eqa:rwpot} to compute the source substructures' weights $\mathbf{w}_s$. Set $\mathbf{w}_t = \frac{\mathbf{1}_{k_t}}{k_t}$
	\State Use GCG to compute the optimal coupling matrix $\boldsymbol{\pi}^*$
	\State According to \equationname~\eqref{eqa:baymap}, compute the transformation of the source substructures and obtain $\hat{\mathbf{P}_s}$
	\State Use $\hat{\mathbf{P}}_s$ and $\tilde{\mathbf{Y}}_s$ to build the model and predict the $\mathbf{P}_t$. The predictions are noted as $\tilde{\mathbf{Y}}_t$
	\State Assign $\tilde{y}_{t,i}$ to the data belonging to $p_{t,i}$ in the target domain
\end{algorithmic}
\end{algorithm}

\section{Experimental Evaluation}
\label{sec:exper}

In this section, we evaluate the performance of \method via extensive experiments on cross-domain activity recognition.
The source code of our \method is at \url{https://github.com/jindongwang/transferlearning/tree/master/code/traditional/sot}.

\subsection{Dataset and Preprocessing}
\label{subs:e-data}

We adopt four common public datasets. \tablename~\ref{tab:datasetinfo} describes the information and the selected data volume of these datasets. 
In the following, we briefly introduce the basic information of each dataset, and more information can be found in their original papers.

UCI daily and sports dataset (DSADS, D)~\citep{barshan2014recognizing} consists of 19 activities collected from 8 subjects wearing body-worn sensors on 5 body parts. UCI human activity recognition using smartphones data set (UCI-HAR, H)~\citep{anguita2012human} is collected by 30 subjects performing 6 daily living activities with a waist-mounted smartphone. USC-SIPI human activity dataset (USC-HAD, U)~\citep{mi12:ubicomp-sagaware} composes of 9 subjects executing 12 activities with a sensor tied on the front right hip. PAMAP2 physical activity monitoring dataset (PAMAP2, P)~\citep{reiss2012introducing} contains data of 18 different physical activities, performed by 9 subjects wearing 3 sensors.

\begin{table}[htbp]
\caption{Information of four datasets. Num means the select data volume. }
\centering
\label{tab:datasetinfo}
\resizebox{\textwidth}{!}{
	\begin{tabular}{c c c  c c c}
		\toprule
		\textbf{Dataset}&\textbf{Subject}&\textbf{Activity}&\textbf{Sample}&\textbf{Location}&\textbf{Num}\\
		\hline
		DSADS&8&19&1.14M&Tarso, Right Arm, Left Arm, Right Leg, Left Leg&2400\\
		UCI-HAR&30&6&1.31M&Waist&6616\\
		USC-HAD&14&12&2.81M&Front Right Hip&4187\\
		PAMAP&9&18&2.84M&Wrist, Chest, Ankle&1688\\
		\bottomrule
	\end{tabular}
}
\end{table}


The cross-domain activity recognition experimental setup is in the following ways. Since different datasets use different sensors and contain different classes of activities, we need to unify our experimental setup for datasets first, and we choose the common parts of the sensors and four common categories of activities. Specifically, we utilize the accelerometer and gyroscope, and each sensor provides 3-axial data (x-, y- and z-axis). We combine them by $\alpha = \sqrt{x^2+ y^2 + z^2}$ following~\citep{wang2018stratified}. Then, according to~\citep{wang2018stratified}, we exploit the sliding window technique to extract features, and 19 features from both time and frequency domains are extracted for a single sensor.
We extract 38 features from one position since two sensors are selected. We choose data from Right Arm, Waist, Front Right Hip, and Right Wrist from these four datasets respectively, and choose four categories, including lying, walking, ascending, descending. In experiments, we perform unsupervised domain adaptation on the target domain.

\subsection{Comparison Methods and Implementation Details}
\label{subs:e-cmid}
We adopt nine comparison methods including both general transfer learning and cross-domain HAR areas:

\begin{enumerate}[noitemsep,nolistsep]
\item Baseline models
\begin{itemize}
	\item 1NN: 1-Nearest Neighbor.
	\item LMNN: Large margin nearest neighbor~\citep{weinberger2009distance}.
\end{itemize}
\item Rough matching
\begin{itemize}
	\item TCA: Transfer component analysis~\citep{pan2010domain}.
	\item SA: Subspace alignment~\citep{fernando2013unsupervised}.
	\item CORAL: CORrelation alignment~\citep{sun2015return}.
	\item STL: Stratified transfer learning~\citep{wang2018stratified}.
\end{itemize}
\item Sample-level matching
\begin{itemize}
	\item OT: Optimal transport~\citep{2013Sinkhorn}.
	\item OTDA: Optimal transport for domain adaptation~\citep{courty2016optimal}.
	\item MLOT: Metric learning optimal transport~\citep{kerdoncuff2020metric}.
\end{itemize}
\end{enumerate}

1NN and LMNN serve as baseline models and TCA, SA, and CORAL perform the rough matching while OT, OTDA, and MLOT belong to sample-level matching methods. And all of these methods use a 1NN classifier for the classification tasks. We conduct experiments in every pair of four datasets and construct 12 tasks in total. 

For most of the comparison methods, we use the codes from \citep{transferlearning.xyz} for implementation. For a particular hyper-parameter configuration, we follow the similar protocol used in~\citep{courty2016optimal}. The target domain is partitioned in validation and test sets. The validation set is used to obtain the best accuracy in the range of the possible hyper-parameters. The hyper-parameter range used follows~\citep{kerdoncuff2020metric} and we slightly reduce the range to fit our task. With the best selected hyper-parameters, we evaluate the performance on the testing set. Classification accuracy on the target domain is adopted as the evaluation metric.

\subsection{Classification Results }\label{subs:e-res}
\begin{table}[t!]
\caption{Activity recognition results on 12 cross-domain tasks.}
\centering
\label{tab:class results}
\resizebox{\textwidth}{!}{
	\begin{tabular}{ c c c c c c c c c c c c c c }
		\toprule
		method&D$\rightarrow$H&D$\rightarrow$U&D$\rightarrow$P&H$\rightarrow$D&H$\rightarrow$U&H$\rightarrow$P&U$\rightarrow$D&U$\rightarrow$H&U$\rightarrow$P&P$\rightarrow$D&P$\rightarrow$H&P$\rightarrow$U&AVG\\
		\hline
		NA&62.70&56.40&66.03&65.16&55.03&60.81&71.30&61.38&60.37&60.99&55.26&51.63&60.59\\
		LMNN&55.24&65.74&48.38&64.27&56.55&65.00&64.58&65.46&67.13&59.53&39.11&41.21&57.68\\
		\hline
		TCA&60.79&51.98&65.66&62.50&41.87&52.06&69.06&53.43&60.88&57.81&46.38&51.45&56.16\\
		SA&63.61&57.36&65.44&66.35&55.60&60.88&70.62&59.64&61.18&62.60&55.45&50.58&60.78\\
		CORAL&64.23&52.25&64.85&64.48&53.03&64.41&68.75&61.93&60.15&60.21&56.20&54.46&60.41\\
		STL&62.83&70.93&65.66&66.15&65.89&67.43&74.69&68.76&65.00&68.96&56.75&55.27&65.69\\
		\hline
		OT&62.13&65.86&65.66&68.91&58.58&67.50&69.90&59.49&63.75&66.77&51.71&57.59&63.15\\
		OTDA&59.36&54.97&65.52&68.91&59.45&67.50&70.26&62.25&63.09&67.19&53.41&59.09&62.58\\
		MLOT&62.53&53.33&64.85&68.12&58.10&62.13&69.53&59.68&61.25&65.99&63.30&49.24&61.51\\
		\hline
		$\mathrm{\method}_c$&\underline{67.74}&\textbf{79.74}&\textbf{73.31}&\textbf{73.39}&\underline{70.87}&\textbf{73.23}&\textbf{80.99}&\textbf{78.04}&\textbf{74.41}&\textbf{76.46}&\textbf{72.82}&\textbf{76.51}&\textbf{74.79}\\
		$\mathrm{\method}_g$&\textbf{74.84}&\textbf{79.74}&\underline{68.68}&\textbf{73.39}&\textbf{71.26}&\textbf{73.23}&\underline{79.90}&\underline{72.82}&\underline{72.28}&\underline{74.48}&\textbf{72.82}&\textbf{76.51}&\underline{74.16}\\
		\bottomrule
	\end{tabular}
}
\end{table}

\begin{figure}[t!]
\centering
\subfigure[OTDA]{
	\label{fig:otda_con}
	\includegraphics[width=0.3\textwidth]{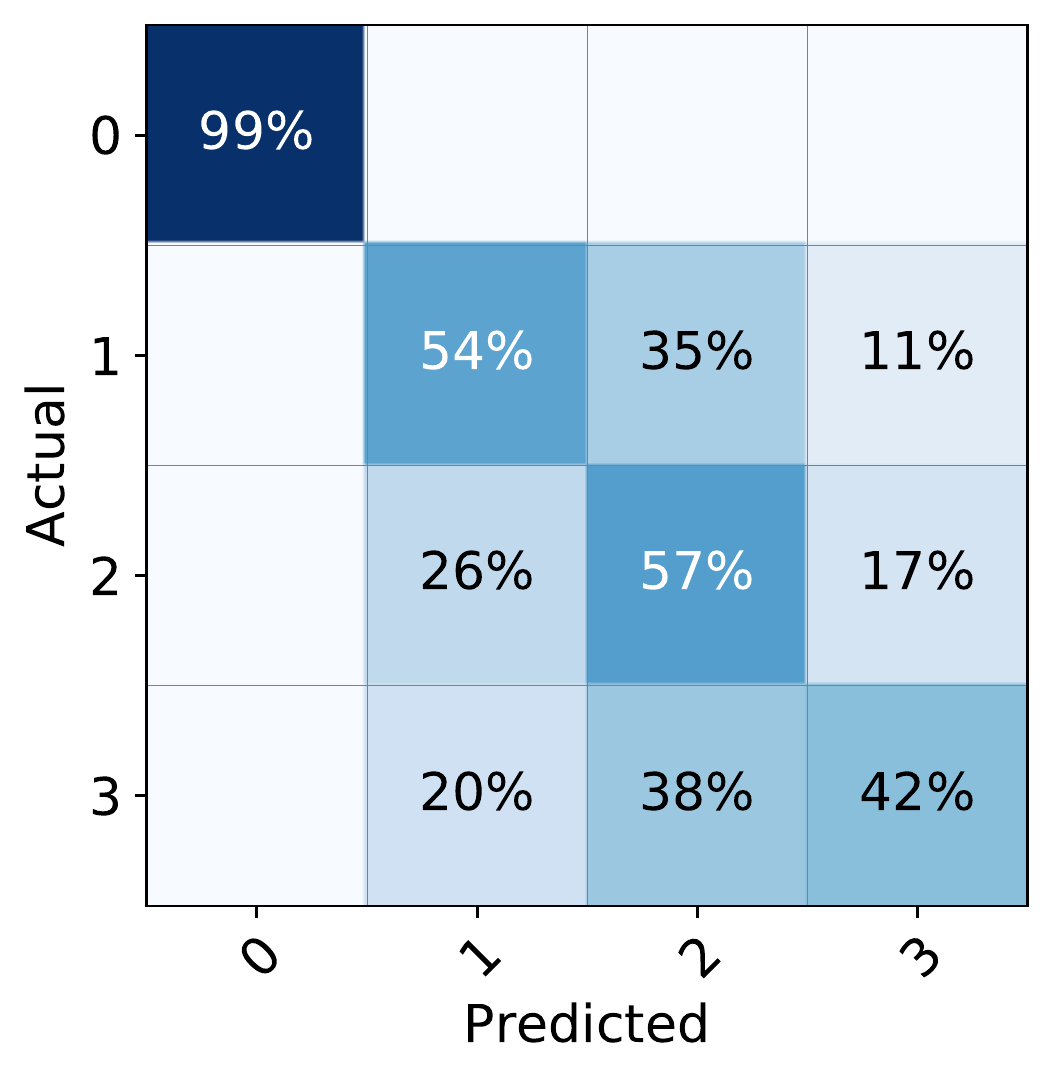}
}
\subfigure[$\mathrm{\method}_c$]{
	\label{fig:gmmot_con}
	\includegraphics[width=0.3\textwidth]{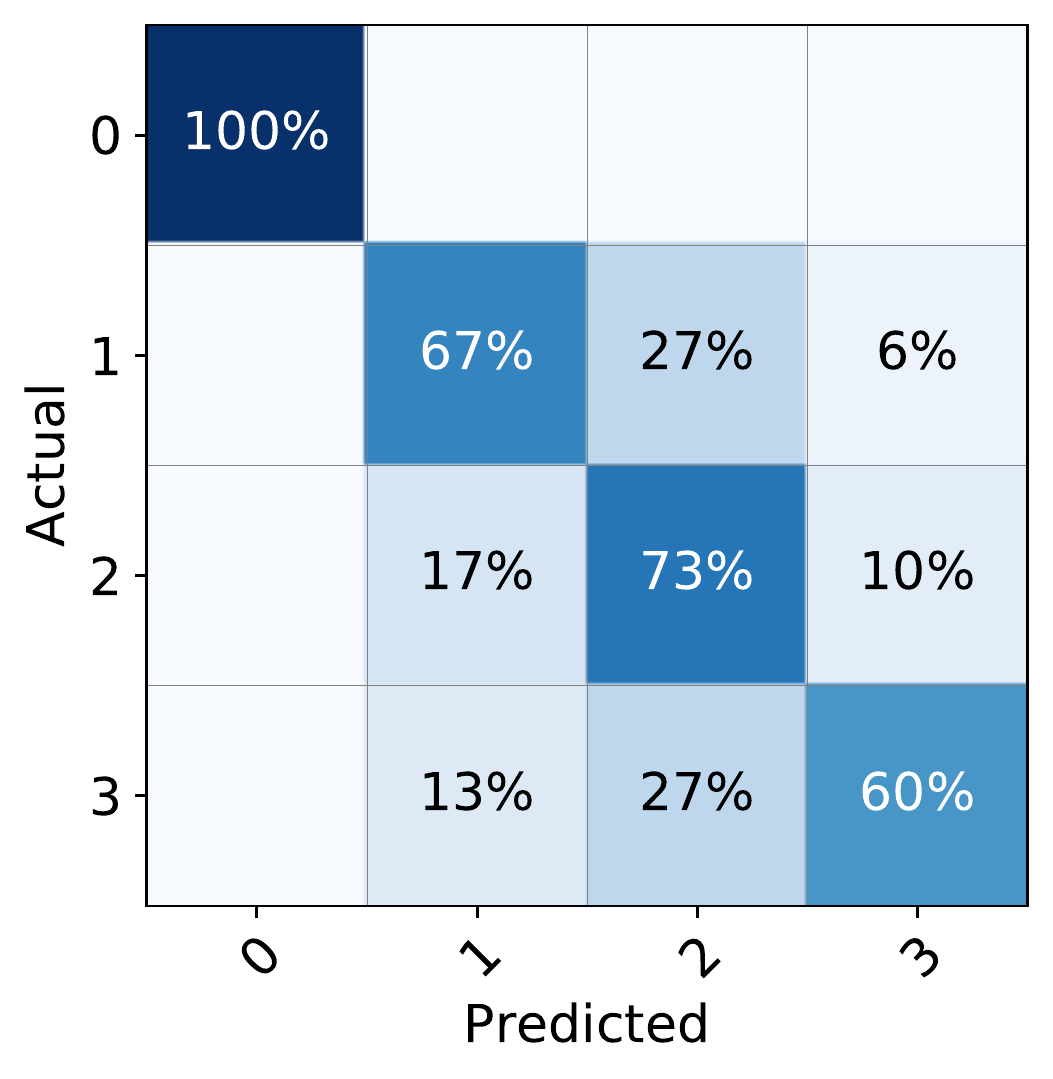}
}
\caption{Confusion matrices for $U\rightarrow D$.}
\label{fig:conm}
\end{figure} 

\begin{figure}[htbp]
\centering
\subfigure[T-SNE for substructures]{
	\label{fig:tsne-gmmot}
	\includegraphics[width=0.3\textwidth]{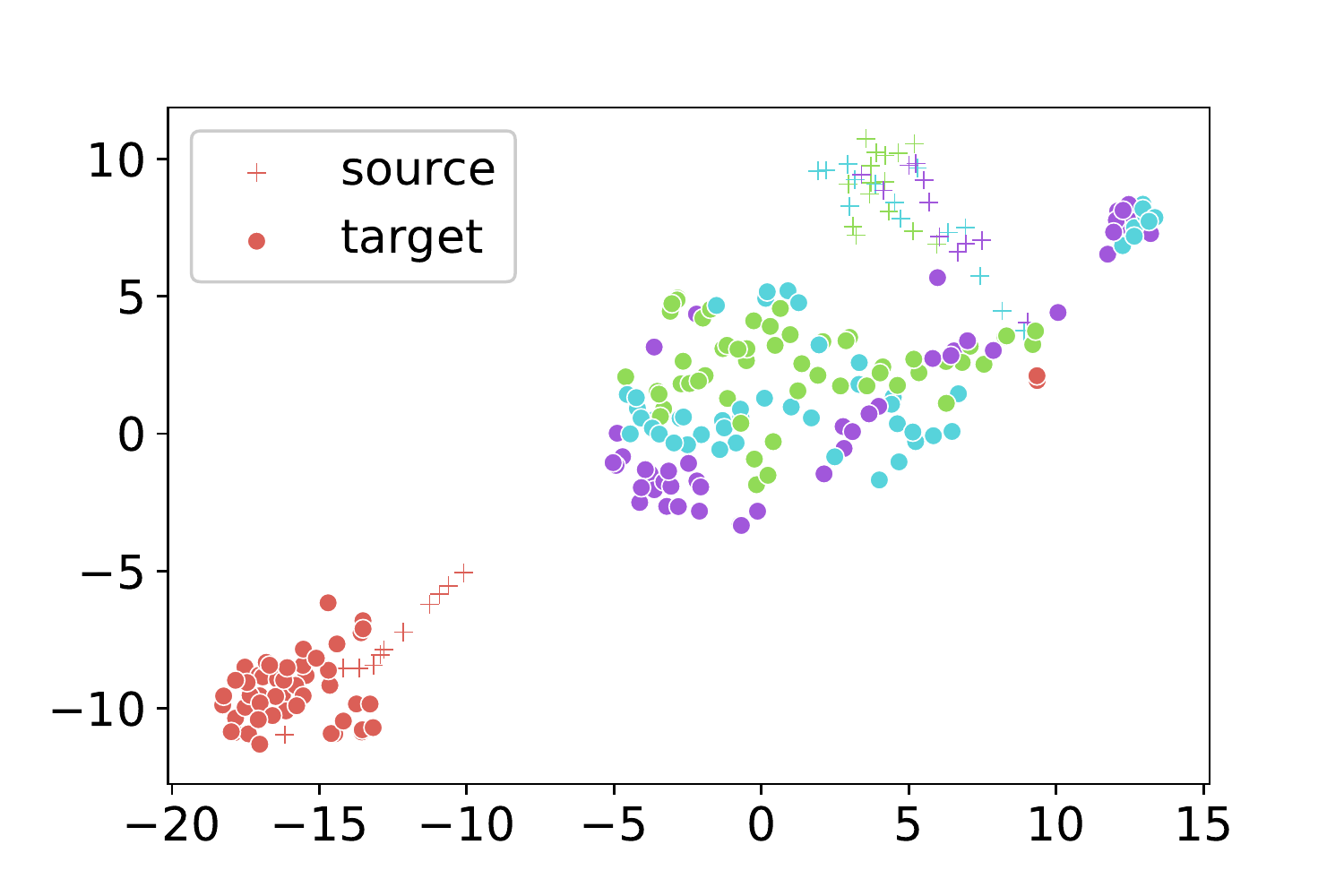}
}
\subfigure[Errors in substructures]{
	\label{fig:gmmot_error}
	\includegraphics[width=0.3\textwidth]{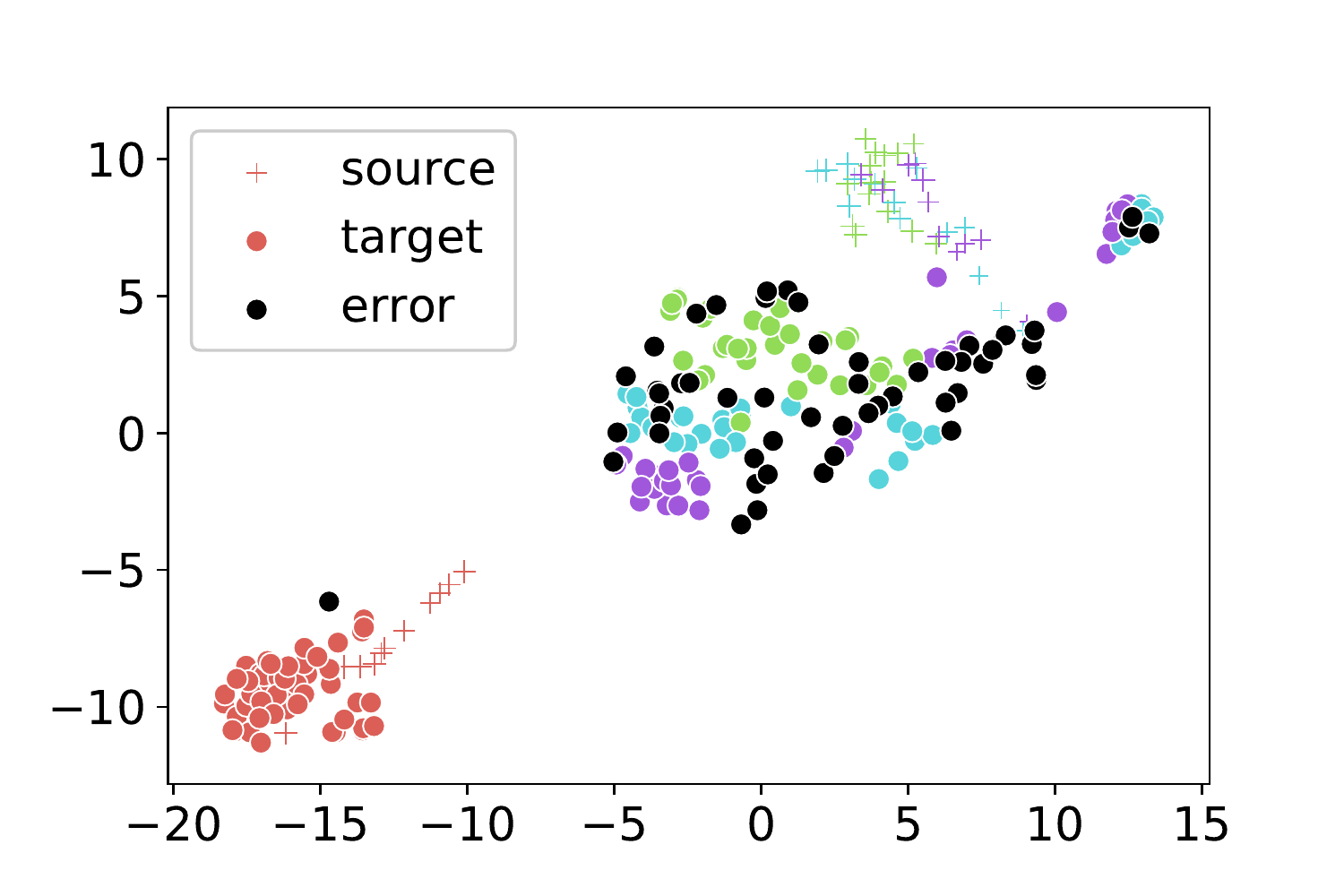}
}
\subfigure[Errors in data]{
	\label{fig:gmmot_error_d}
	\includegraphics[width=0.3\textwidth]{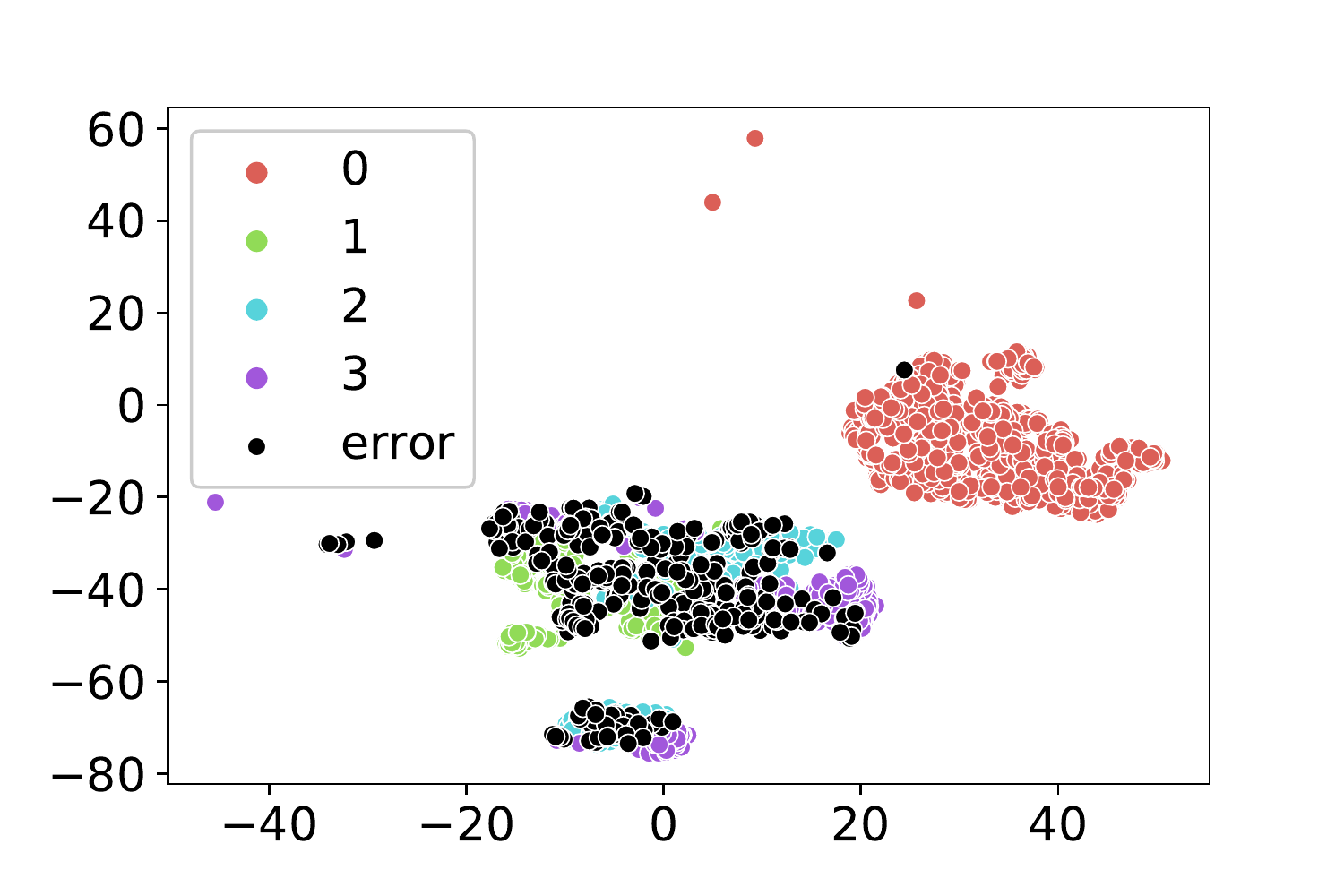}
}
\caption{Visualization for $U\rightarrow D$ using \method. Different colors mean different classes while black means error predictions. The boundaries of different colors correspond to the boundaries of different classes.}
\label{fig:gmmotsne}
\end{figure}

The results of classification are shown in \tablename~\ref{tab:class results} and \figurename~\ref{fig:conm}. From these results, we have the following observations: 1) Both $\mathrm{\method}_c$ and $\mathrm{\method}_g$ achieve the best classification accuracy on all tasks. It is obvious that \method significantly outperforms other methods with a remarkable improvement (over $\mathbf{9}\%$ on average). 2) Compared to baseline methods, rough matching methods and sample-level matching methods only have slight improvements due to neglecting the details or introducing much noise. Thus, being too rough or too delicate is not suitable for cross-dataset activity recognition. 3) \figurename~\ref{fig:conm} shows lying is easy to identify correctly while walking, ascending and descending are difficult to classify, which is in line with the intuition and is consistent with \figurename~\ref{fig:tsne-gmmot}. 4) $\mathrm{\method}_g$ is slightly worse than $\mathrm{\method}_c$. Maybe because $\mathrm{\method}_g$ uses more approximations when computing. In the following experiments, we use $\mathrm{\method}_c$ by default if there is no special instruction.

\subsection{Visualization Study}
\label{subs:e-vis}
As we can see in \figurename~\ref{fig:tsne-gmmot}, the number of points is much smaller after clustering, which can bring high efficiency, and the margins between points are bigger.
The class of the substructures in the target domain is temporarily determined by most of the data in the corresponding cluster.
In \figurename~\ref{fig:gmmot_error}, we can see that it is easy to misclassify the points which are near the margins or intersected with other classes' points. Due to the bigger margins and the fewer intersecting points, the accuracy on the substructures is $73\%$ while the accuracy on raw data using OTDA is $70.375\%$. \figurename~\ref{fig:gmmot_error_d} shows the misclassified data using $\mathrm{\method}_c$ and the accuracy is improved over $\mathbf{9}\%$ due to the exploitation of the substructures.

\subsection{Ablation Study}\label{subs:e-abla}
We first demonstrate \methodda is not limited to its OT implementation (i.e., SOT), but a general framework, and then evaluate the importance and robustness of three important parts of \method, namely, substructures generation, weighting source substructures and OT-based mapping of substructures.
\begin{figure}[htbp]
\centering
\subfigure[CORAL]{
	\label{fig:coralm}
	\includegraphics[width=0.4\textwidth]{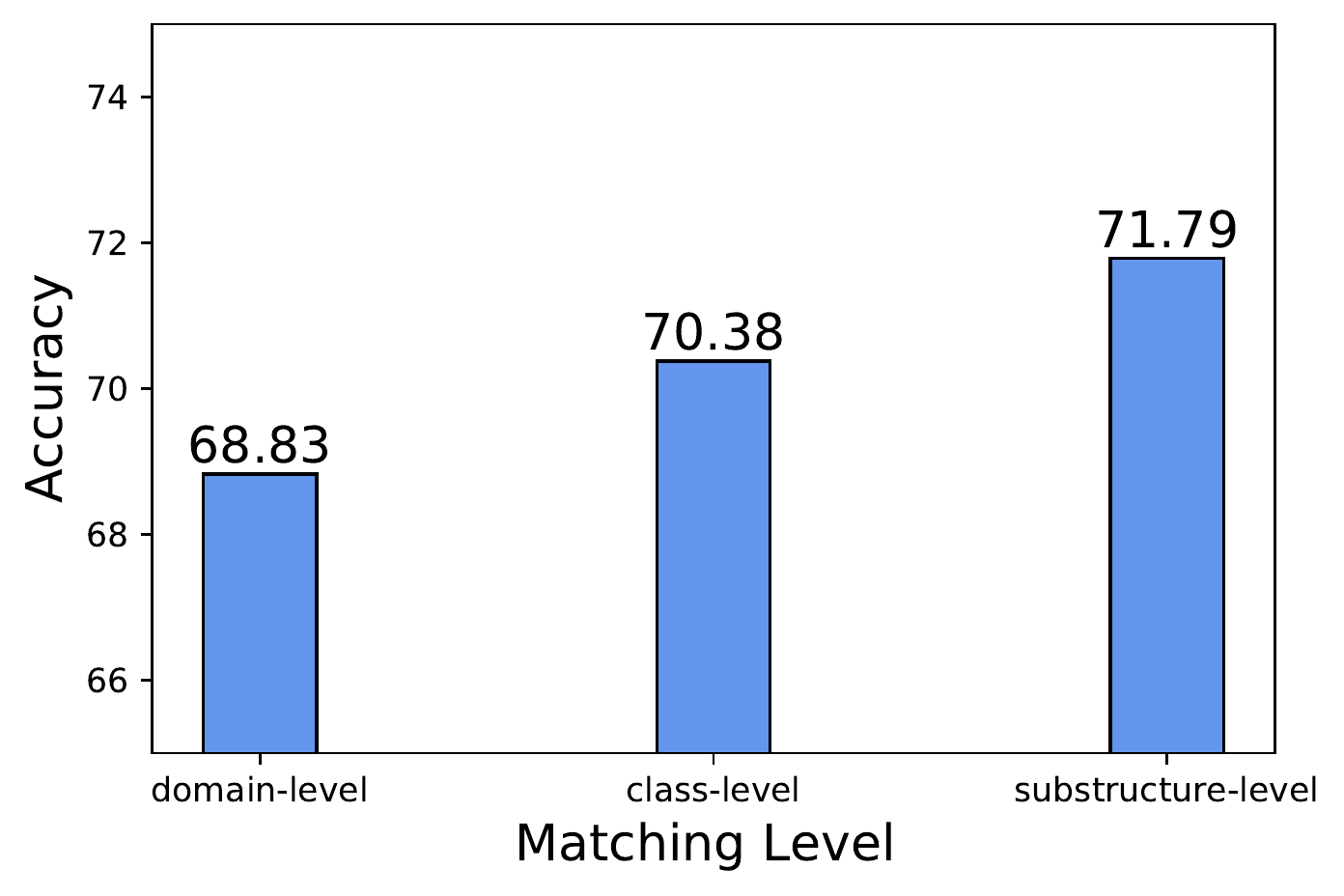}
}
\subfigure[OTDA]{
	\label{fig:otdam}
	\includegraphics[width=0.4\textwidth]{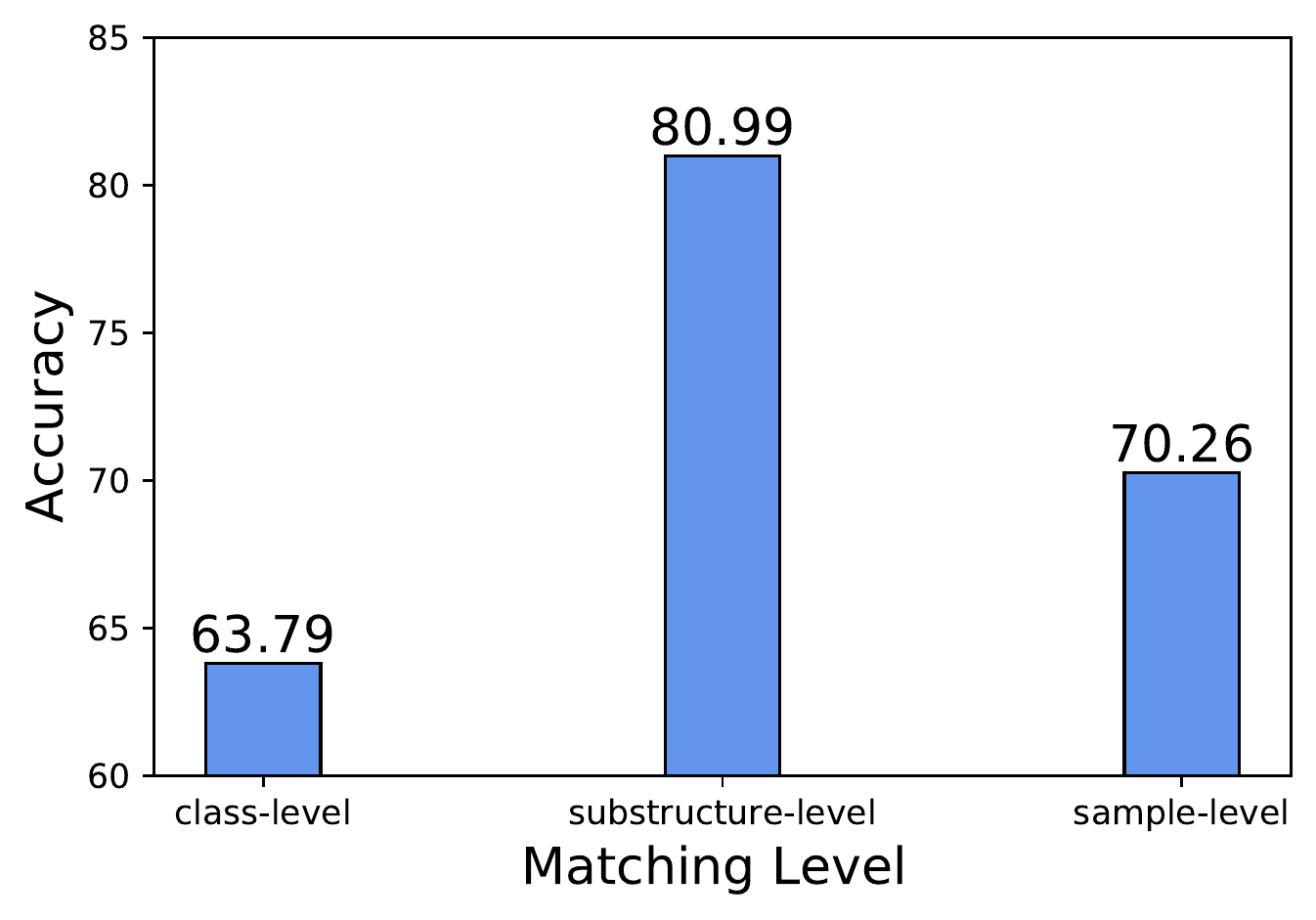}
}
\caption{Accuracy of different matching levels with different methods for task $U\rightarrow D$.}
\label{fig:SSDA}
\end{figure}

\subsubsection{Implementing \methodda using other methods}
\label{subs-ab-ssda}

We implement CORAL with domain-, class- and substructure-level matching and \figurename~\ref{fig:coralm} shows that class-level matching CORAL gets better accuracy than domain-level matching while substructure-level matching CORAL achieves the best accuracy, which demonstrates finer matching gets better results. In addition, we implement OTDA with class-, substructure- and sample-level matching, and \figurename~\ref{fig:otdam} illustrates that substructure-level matching OTDA gets a better accuracy than the sample-level matching OTDA, which may be substructure-level matching is robust to noise. Overall, \figurename~\ref{fig:SSDA} demonstrates \methodda is a general framework and can achieve commendable results.

\begin{figure}[htbp]
\centering
\subfigure[Results for $U\rightarrow D$ using different random params for GMM]{
	\label{fig:gmmotabld-c}
	\includegraphics[width=0.4\textwidth]{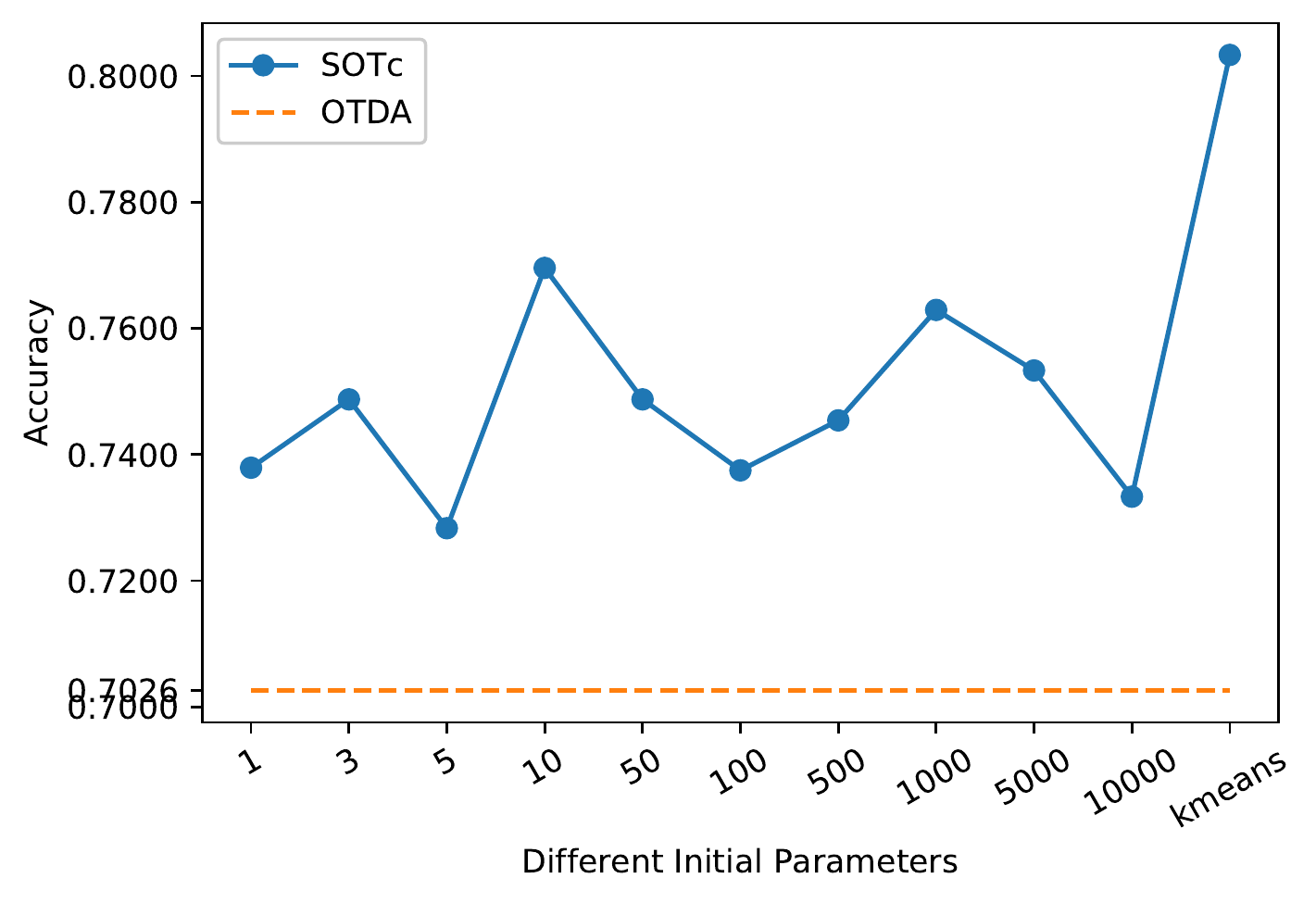}
}
\subfigure[Results for $U\rightarrow D$ using substructures with uniform or different weights]{
	\label{fig:wgmmot}
	\includegraphics[width=0.4\textwidth]{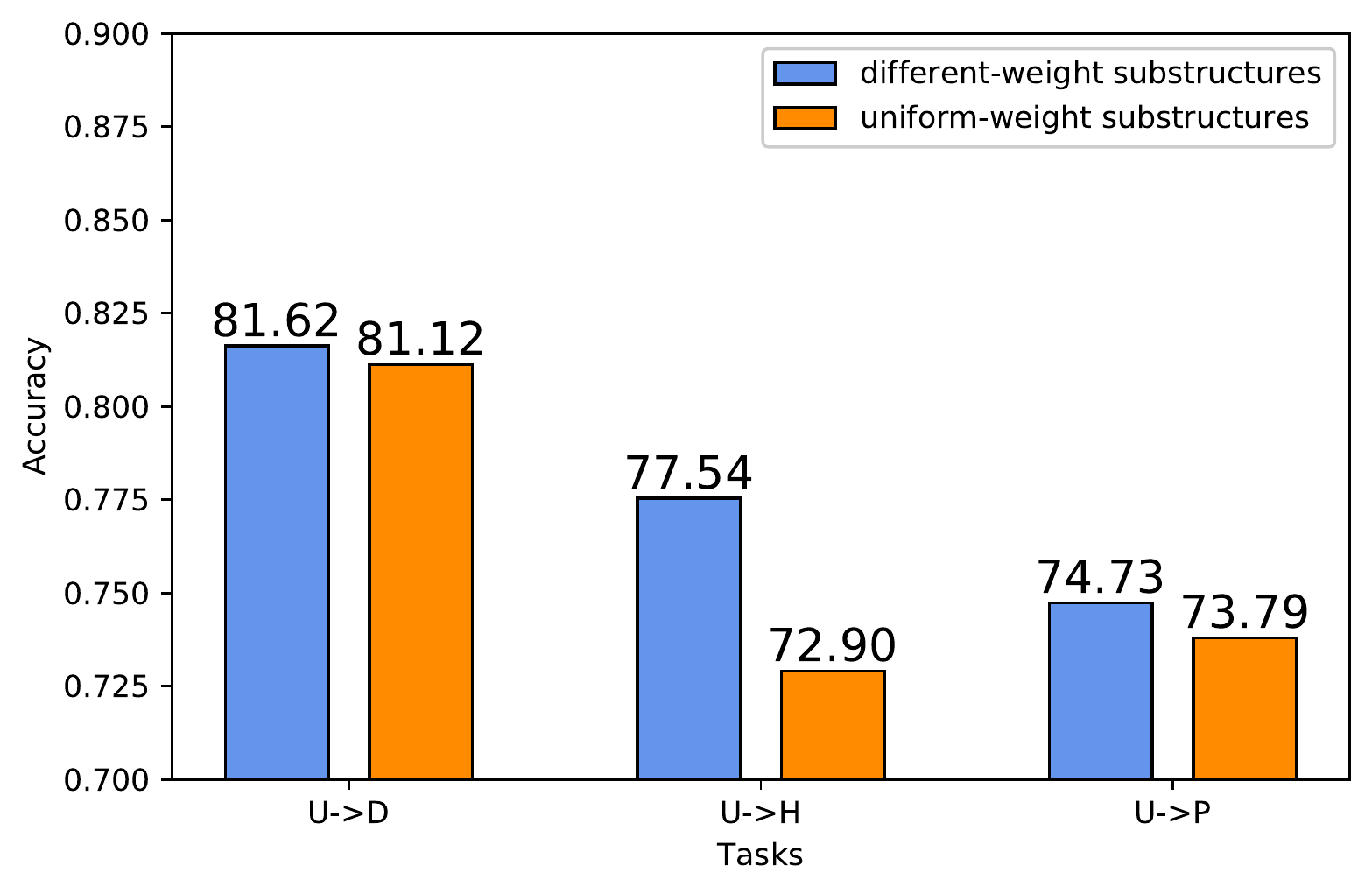}
}
\caption{Ablation study of substructures generation and weighting source substructures.}
\label{fig:SOT2-3abl}
\end{figure}

\subsubsection{Substructures generation}
\label{subs-ab-sg}

We generate substructures with GMM using different initial parameters and the results are in \figurename~\ref{fig:gmmotabld-c}.
The x-axis shows the different initial states. Obviously, \method performs better than OTDA on any random initial states. The initial parameters obtained from k-means get the best accuracy, which indicates that good clustering results bring high accuracy.

\subsubsection{Weighting source substructures}
\label{subs-ab-wss}
To demonstrate the effect of weighting source substructures, we compare the accuracy between the experiments with it and without it. Without weighting source substructures, we simply assign the same weight to all the source substructures. \figurename~\ref{fig:wgmmot} shows there is an improvement with weighting source substructures.
\begin{figure}[htbp]
\centering
\subfigure[$\eta = 0$]{
	\label{fig:noeta}
	\includegraphics[width=0.4\textwidth]{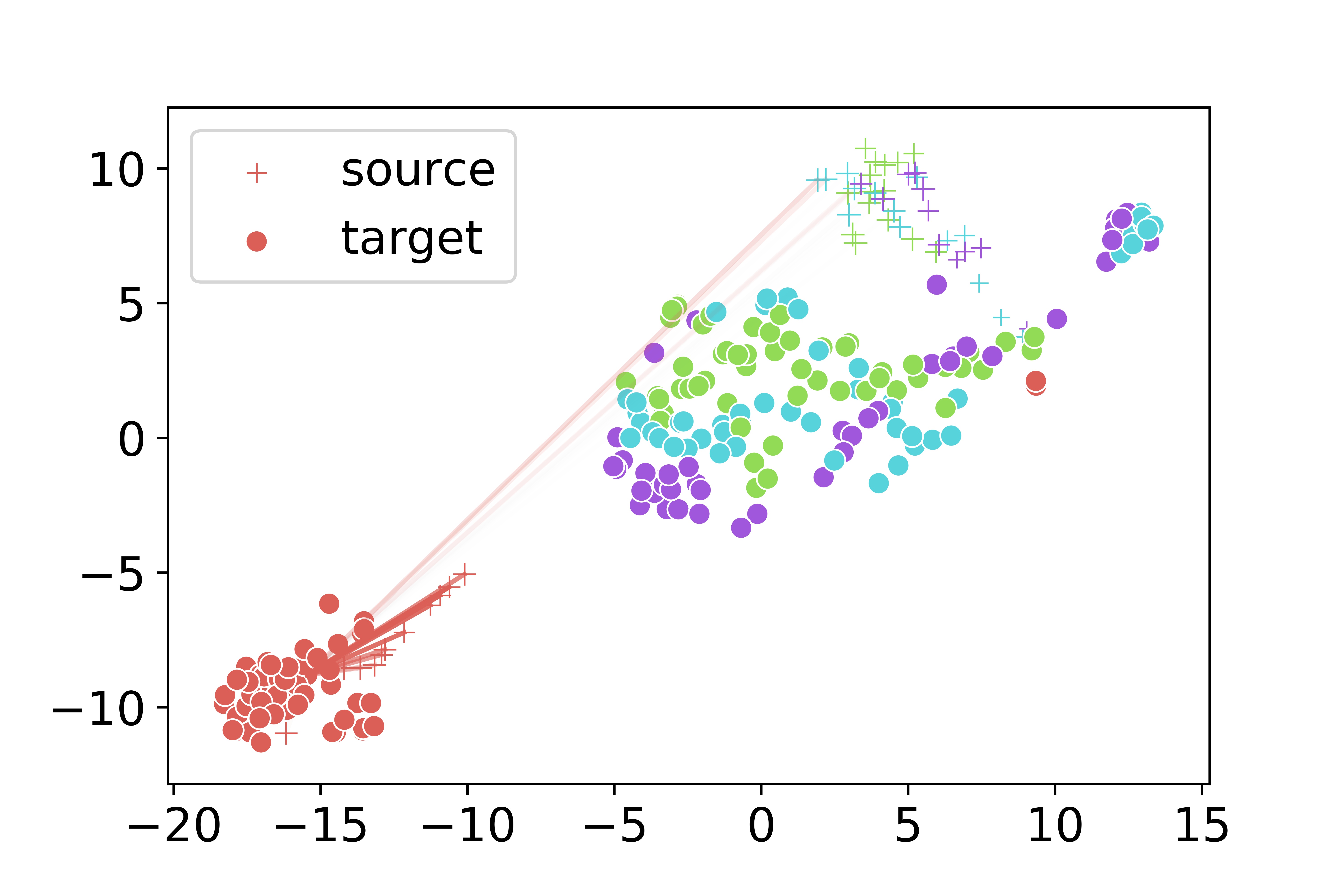}
}
\subfigure[$\eta = 0.5$]{
	\label{fig:eta05}
	\includegraphics[width=0.4\textwidth]{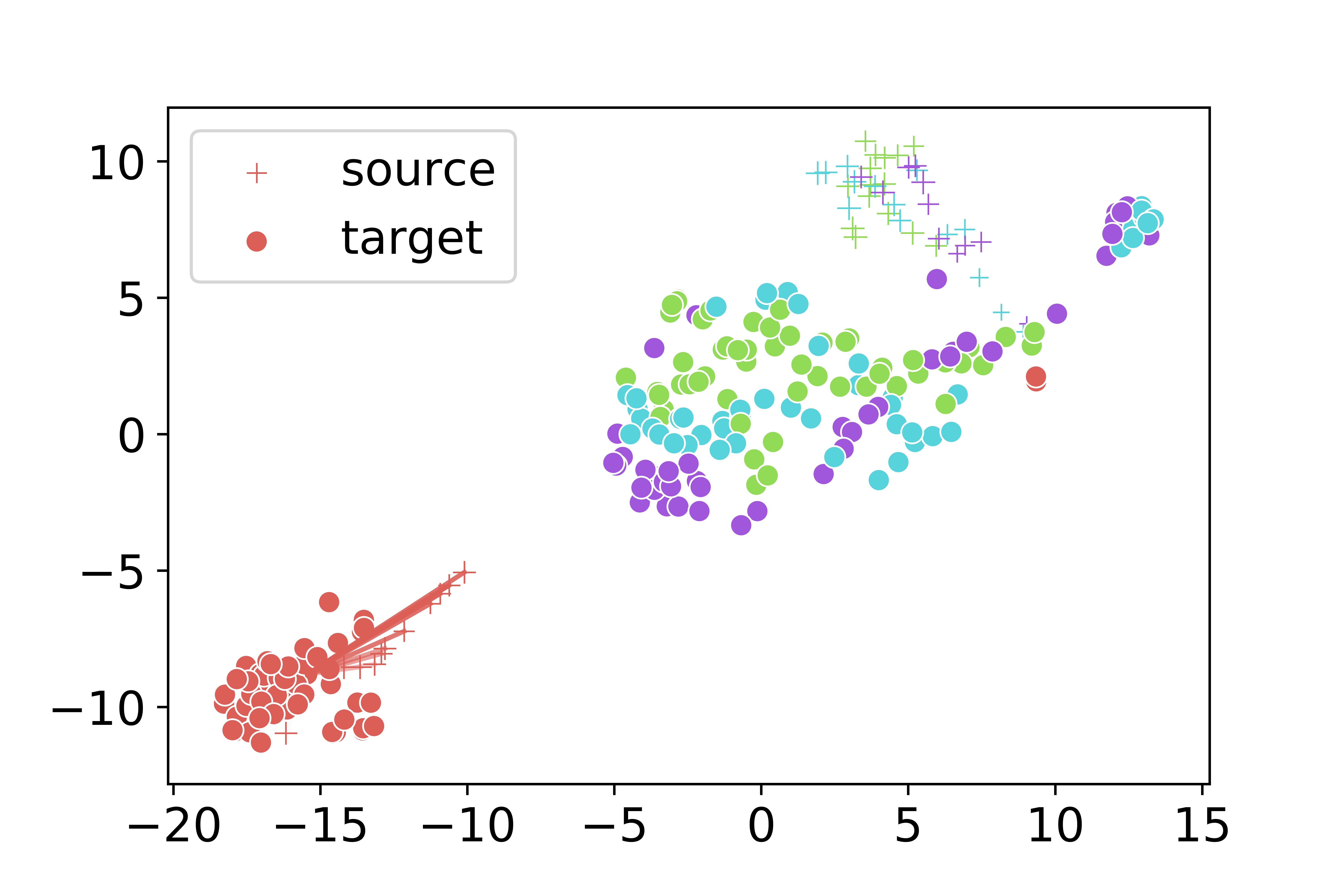}
}
\caption{The group regularizer's function.}
\label{fig:groupreg}
\end{figure}

\subsubsection{OT-based mapping of substructures}~\label{subs-ab-ots}
In this part, we illustrate the function of the group regularizer. No group regularizer is used in \figurename~\ref{fig:noeta} while there is a group regularizer with $\eta = 0.5$ is used in \figurename~\ref{fig:eta05}. In \figurename~\ref{fig:noeta}, some points with different colors are linked with the red point while only red points are linked with the red point in \figurename~\ref{fig:eta05}, which are caused by the group regularizer.

\subsection{Parameter Sensitivity}\label{subs:e-parasen}
\begin{figure}[htbp]
\centering
\subfigure[$\lambda_1$]{
	\label{fig:regce}
	\includegraphics[width=0.21\textwidth]{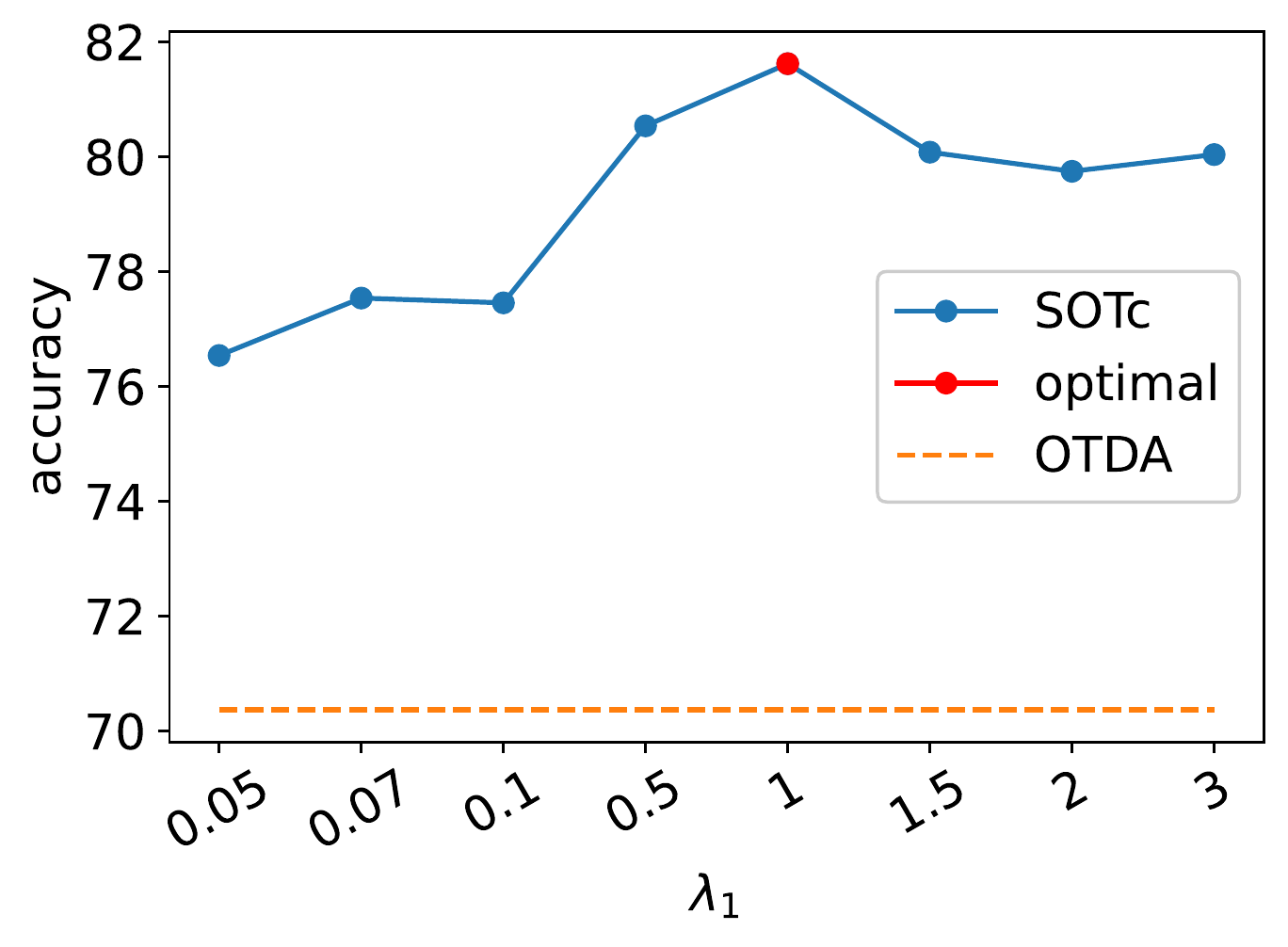}
}
\subfigure[$\lambda$]{
	\label{fig:rege}
	\includegraphics[width=0.21\textwidth]{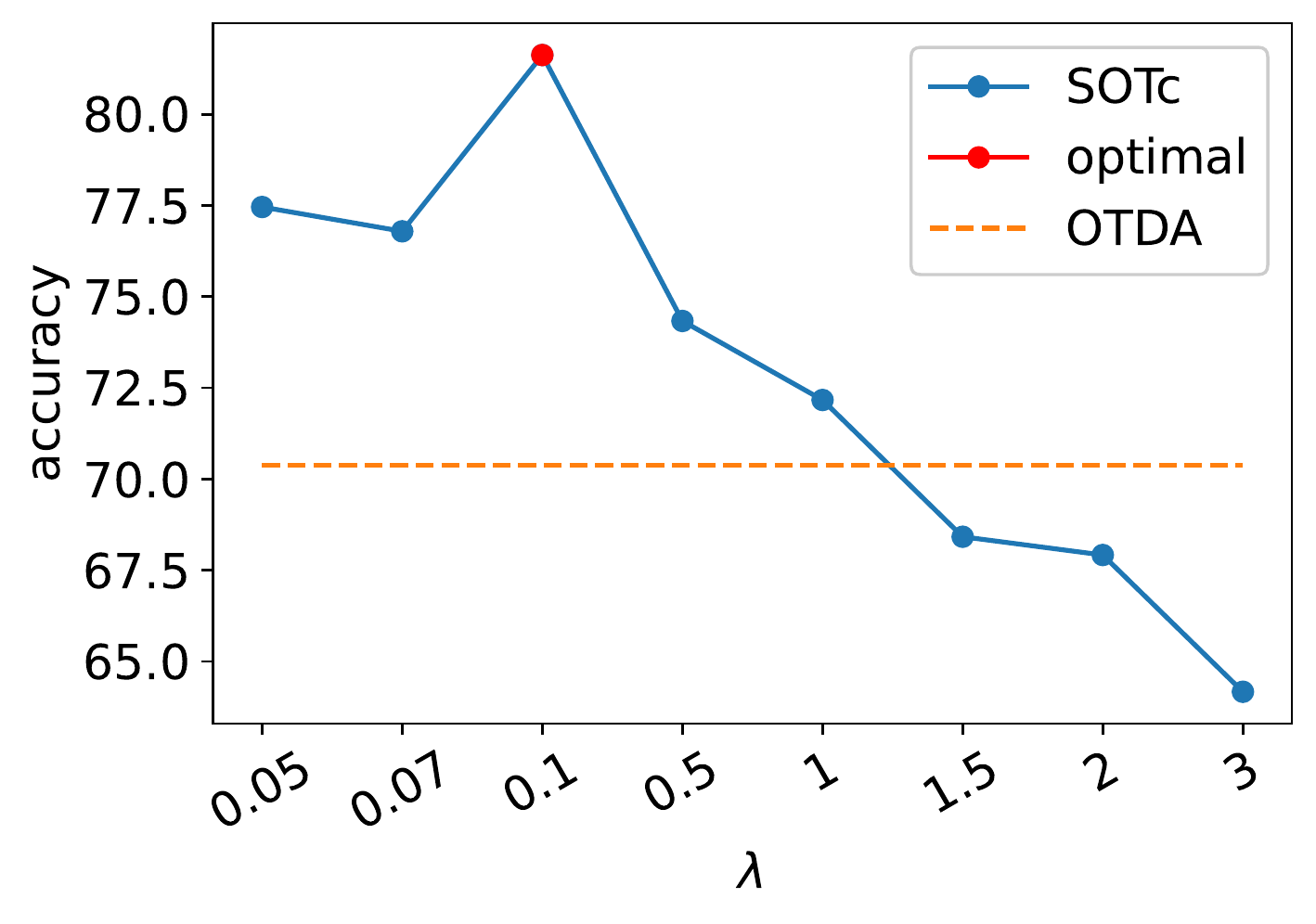}
}
\subfigure[$\eta$]{
	\label{fig:eta}
	\includegraphics[width=0.21\textwidth]{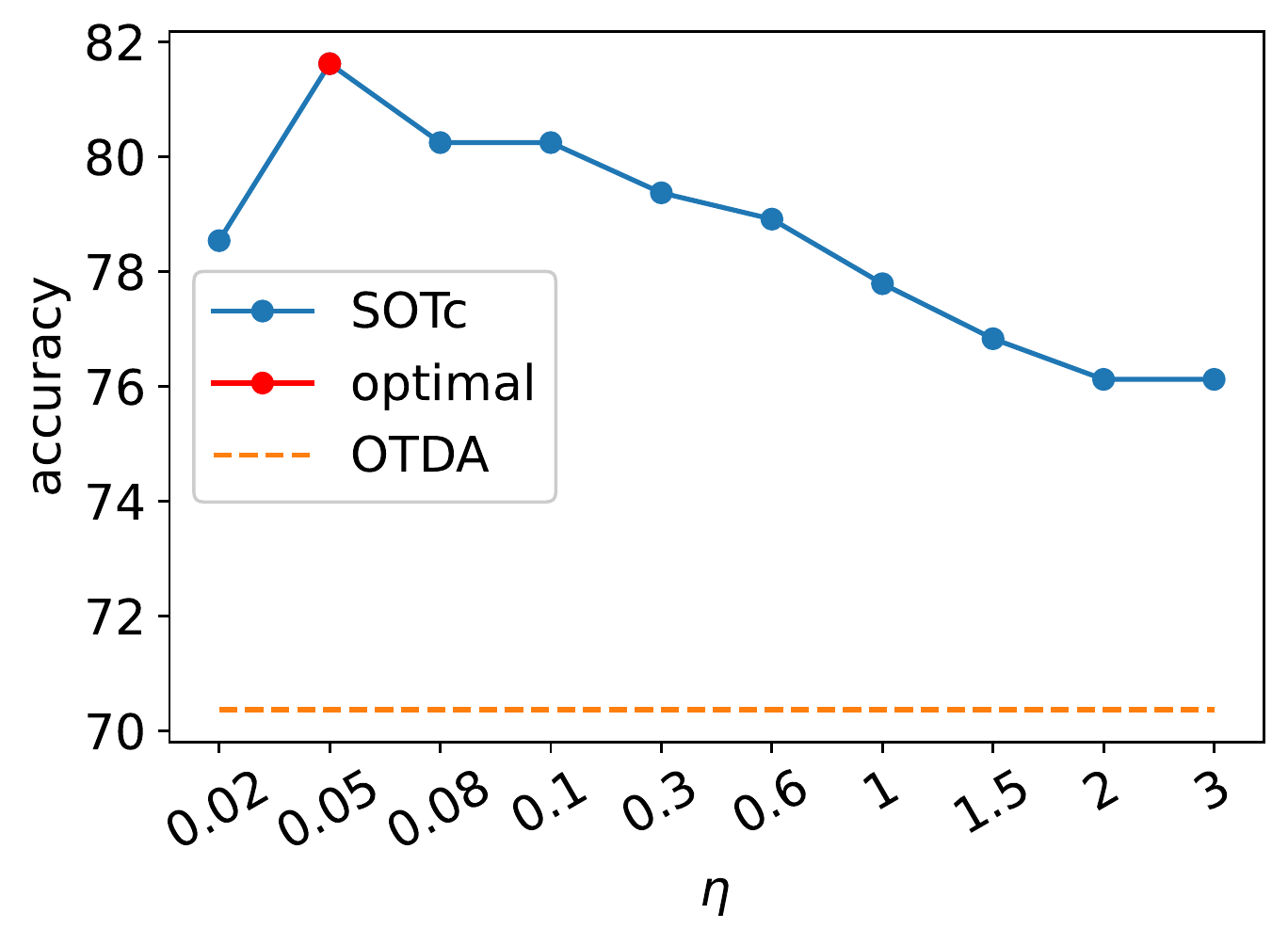}
}
\subfigure[$k_t$]{
	\label{fig:d}
	\includegraphics[width=0.21\textwidth]{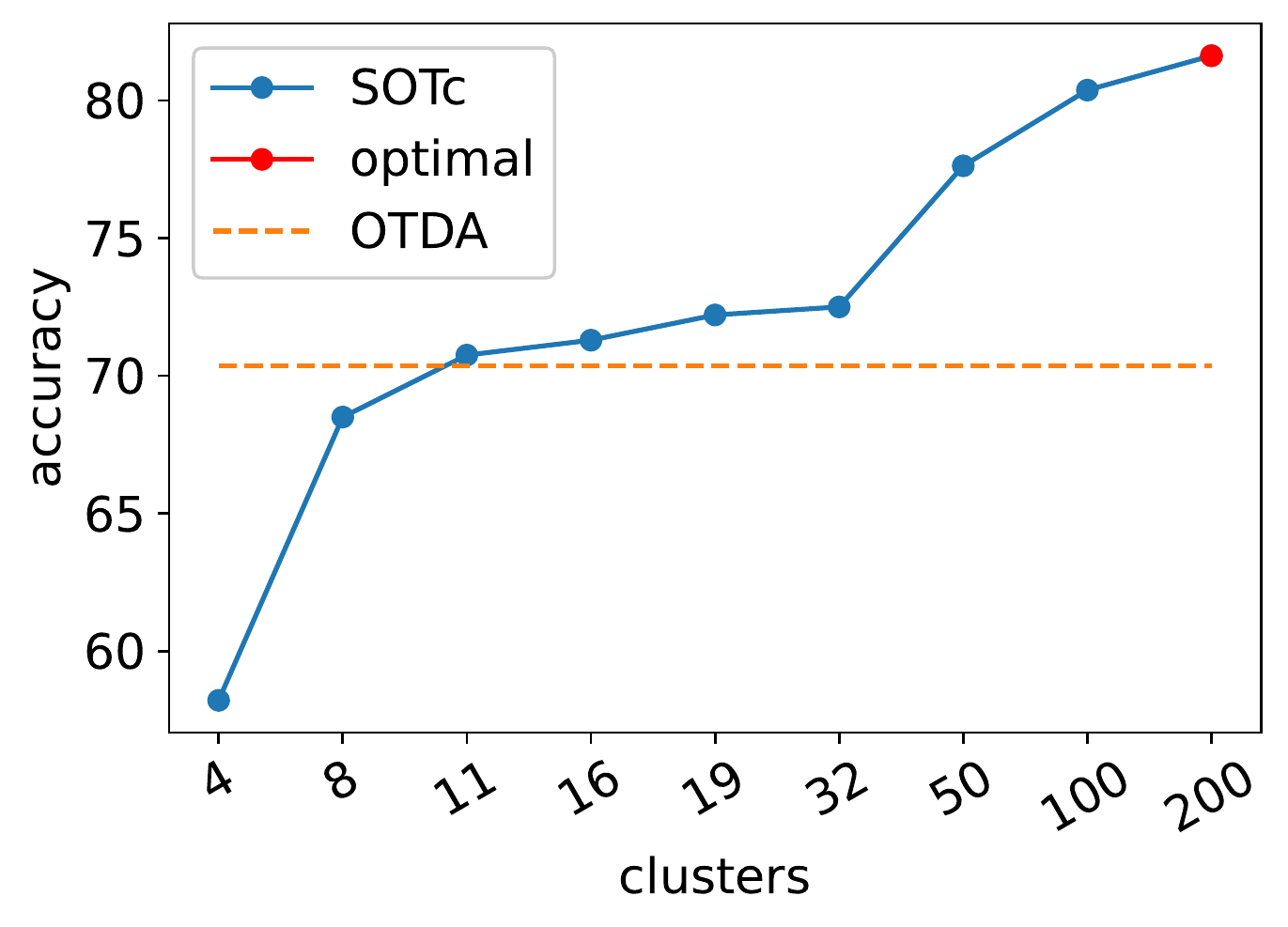}
}
\caption{Influence of different hyper-parameters.}
\label{fig:paramsen}
\end{figure}

In this section, we evaluate the parameter sensitivity of \method. \method involves four parameters: $\lambda_1$, $\lambda$, $\eta$ and $k_t$. We change one parameter and fix the other parameters to observe the performance of \method. In \figurename~\ref{fig:paramsen}, the red points are the optimal points, and we observe the surrounding results. From \ref{fig:regce}-\ref{fig:d}, we can see that the results with parameters around the red points are all better than OTDA. The results reveal that \method is more effective and robust than other methods under different parameters near the optimal.  
\subsection{Convergence and Time Complexity}\label{subs:e-conver}

\begin{figure}[htbp]
\centering
\subfigure[Convergence]{
	\label{fig:loss}
	\includegraphics[width=0.3\textwidth]{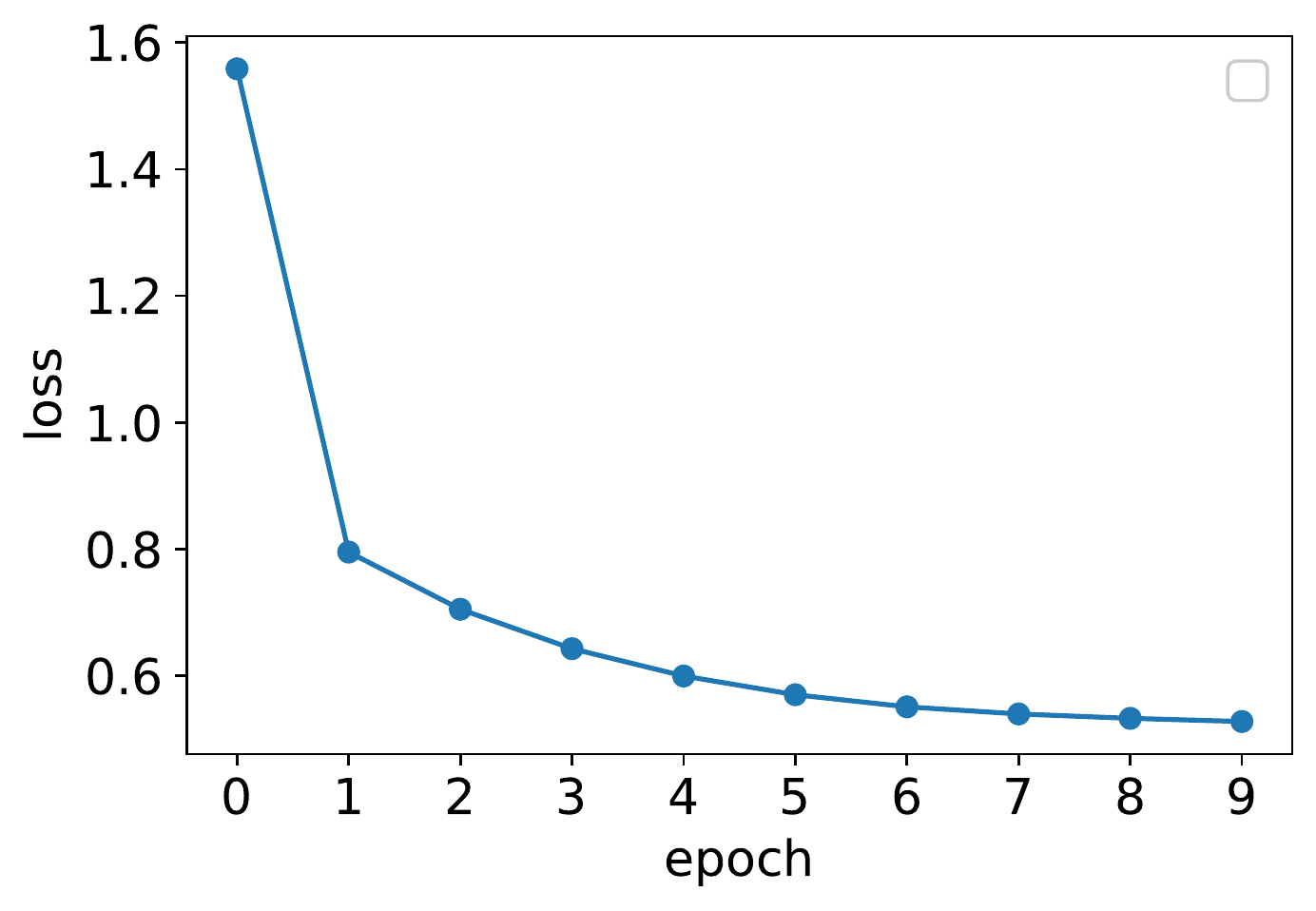}
}
\subfigure[Time Complexity and Accuracy Comparisons]{
	\label{fig:Tcompar}
	\includegraphics[width=0.3\textwidth]{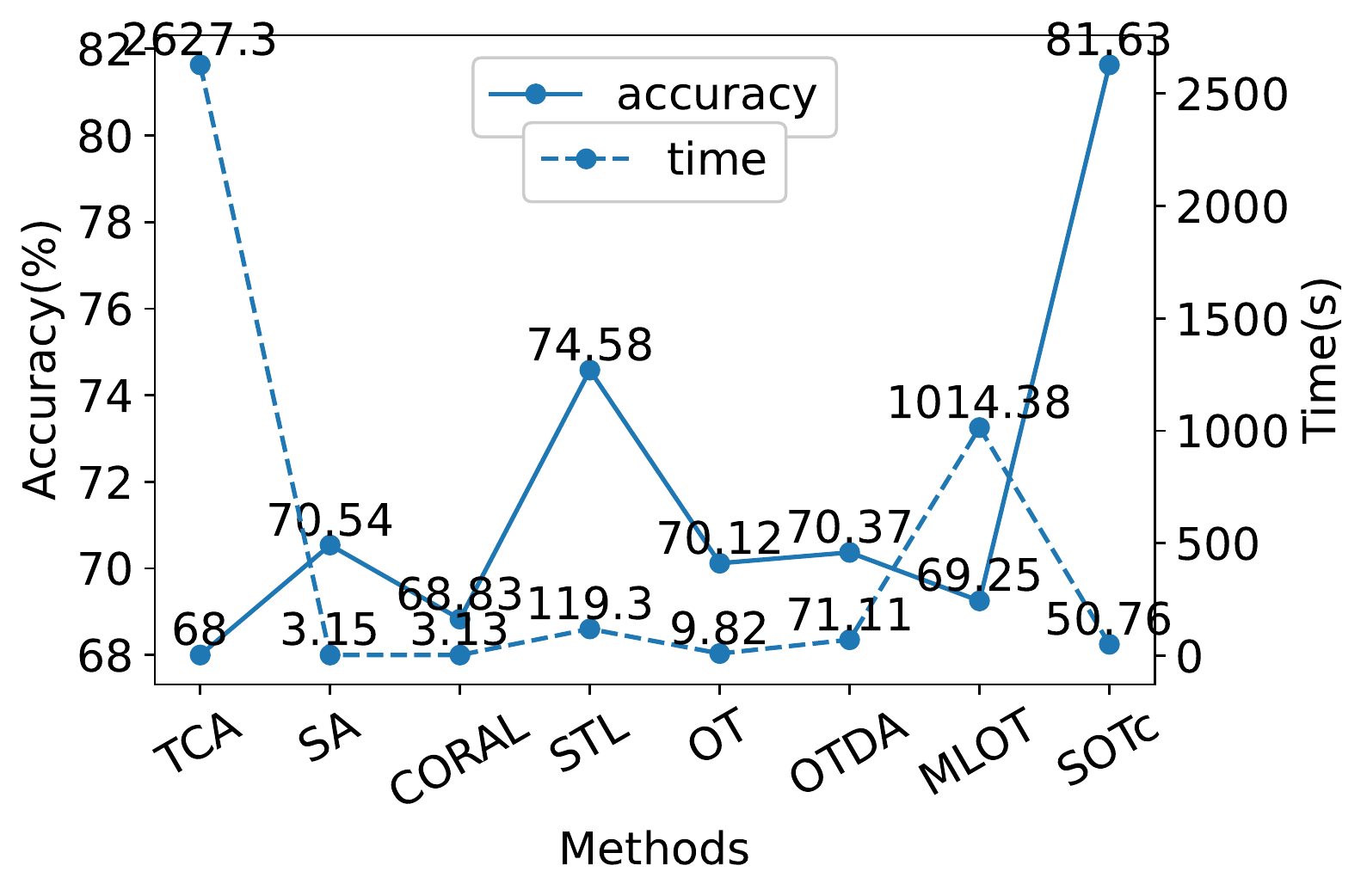}
}
\subfigure[Time Spent of Each Part in \method]{
	\label{fig:Tpart}
	\includegraphics[width=0.3\textwidth]{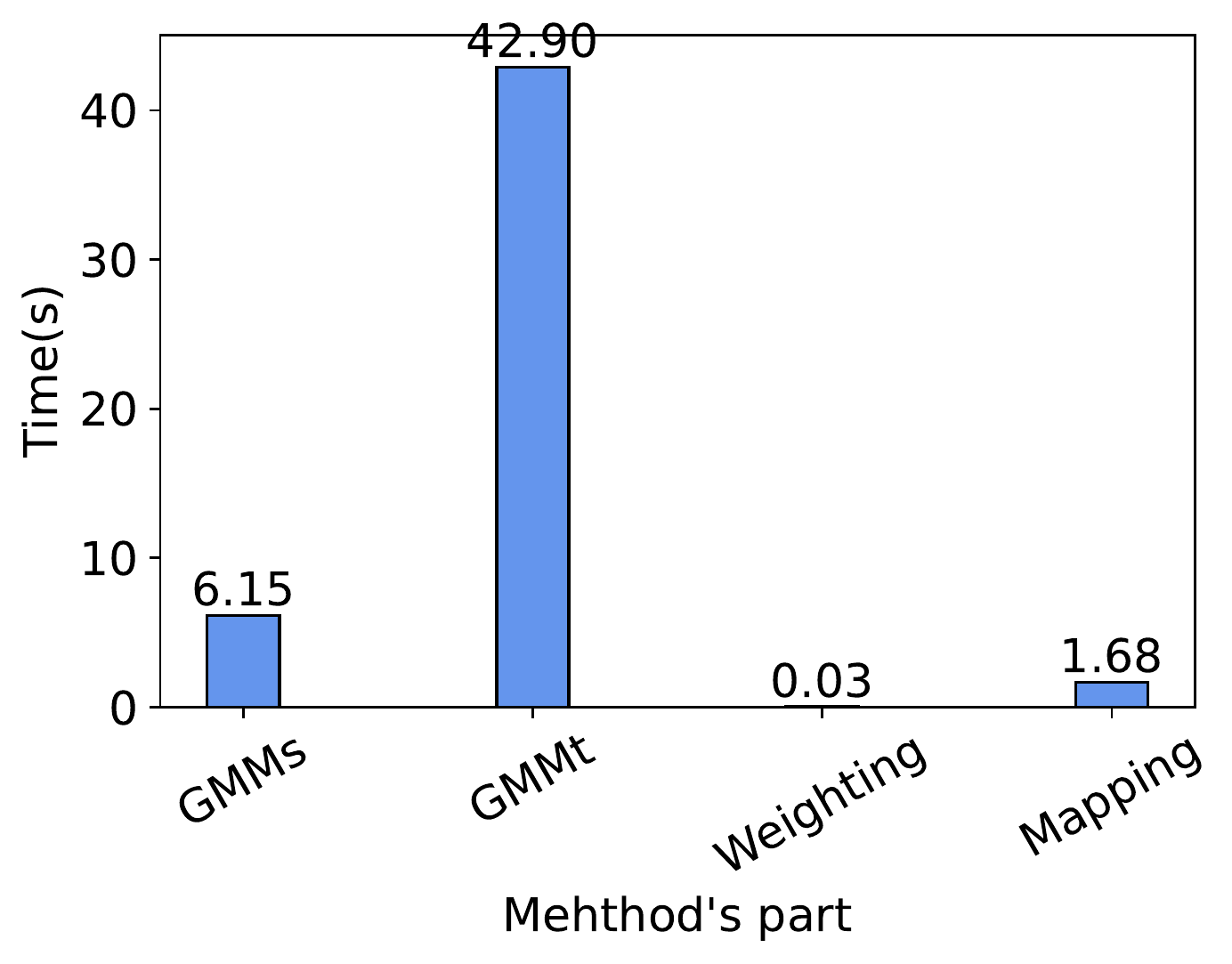}
}
\caption{Convergence and Time Complexity.}
\label{fig:CandT}
\end{figure}

In this section, we investigate the convergence and time complexity. In \figurename~\ref{fig:loss}, we can see \method convergences in the 10th epoch. And in the actual experiments, 20 epochs are enough for \method. This means our \method can reach fast convergence. With the same hyper-parameters, \method is $\mathbf{5} \times$ faster than traditional OT-based DA methods.

To compare the time complexity, we conduct each experiment 10 times and sum over the time. \figurename~\ref{fig:Tcompar} indicates that \method gets the best accuracy while the time spent is much less than TCA and MLOT. When we slightly change the parameters of OTDA, the time used changes from $71.11$s to $285.32$s. From \figurename~\ref{fig:Tpart}, we can see that GMMs use most of the time in \method. These two parts can be fixed in real experiments, which means we only need to pay attention to the parts of weighting and mapping. Obviously, the time used in the weighting part is negligible, and the time used in the mapping part is smaller than all the other methods compared. In the following, we analyze the time complexity theoretically. 

As we know, without the entropy regularizer, combinatorial algorithms, such as the simplex methods and its network variants, are used to solve the optimal transport. However, their computational complexity is shown to be $O((n_s + n_t)n_sn_t \log(n_s + n_t))$, which is impossible to handle large datasets. When handling the OT with the entropy regularizer, we can get an algorithm with the quadratic complexity, which is still below our expectations. The key problem is that $n_s$ and $n_t$ are too large to get the fast results. To deal with this problem, we use the substructures instead of initial data points. The numbers of the substructures are $k_s$ and $k_t$ respectively, and they are much smaller than $n_s$ and $n_t$. To get the substructures, GMM, whose time complexity is $O(L_1KN)$, is adopted. $K$ is the number of clusters, $N$ is the number of data, while $L_1$ is the number of iterations. We can get the weights of the substructures with one matrix operation and the time spent on this operation is negligible. To sum up, the time complexity of our method is about $O(L_1KN + L_2K^2)$, where $L_1$ and $L_2$ are the numbers of the iterations, and they are much smaller than N.

\section{Conclusions and Future Work}\label{sec:concl}
Leveraging labeled data from auxiliary domains is a usual way to deal with the label scarcity problem in Human Activity Recognition. In this paper, we propose \methodda to utilize substructures and propose an OT based implementation, \method, for cross-domain activity recognition. Comparing to existing methods which perform rough matching or sample-level matching, \methodda obtains the internal substructures and completes substructures-level matching which considers more fine-grained locality information of domains and is robust to noise in a certain degree. Comprehensive experiments on four large public datasets demonstrate the significant superiority of \method over other state-of-the-art methods. In addition, \method is much faster than other methods, which means it can be used in the larger datasets.

In the future, we plan to extend \method using deep clustering, as well as applying \methodda to other fine-grained activity recognition problems.

{\small
\bibliography{bibliography}
}

\bibliographystyle{iclr2020_conference}
\end{document}